\documentclass{article}
\usepackage{graphicx}
\usepackage{subcaption}
\usepackage{mathtools}
\usepackage{stmaryrd}
\usepackage{amssymb}
\usepackage{amsthm}
\usepackage{nccmath}
\usepackage{hyperref}
\usepackage{bbm}
\usepackage{appendix}

\usepackage{fullpage}
\usepackage{parskip}
\usepackage{xcolor}

\usepackage{natbib}
\usepackage{apalike}
\usepackage[para]{footmisc}
\usepackage{mathrsfs}

\usepackage{booktabs}
\usepackage{algorithm}
\usepackage{algpseudocode}

\newtheorem{proposition}{Proposition}
\newtheorem{corollary}{Corollary}

\title{Optimal execution on Uniswap v2/v3 under transient price impact}
\author{Bastien Baude\footnote{\href{mailto:bastien.baude@centralesupelec.fr}{bastien.baude@centralesupelec.fr}} \quad Damien Challet\footnote{\href{mailto:damien.challet@centralesupelec.fr}{damien.challet@centralesupelec.fr}} \quad Ioane Muni Toke\footnote{\href{mailto:ioane.muni-toke@centralesupelec.fr}{ioane.muni-toke@centralesupelec.fr}} \\ \\ \textit{Université Paris-Saclay, CentraleSupélec, Laboratoire MICS} \\
\textit{91192 Gif-sur-Yvette, France}}

\begin{document}

\bibliographystyle{apalike}

\maketitle

\begin{abstract}
\setlength{\baselineskip}{10.5pt}
\noindent
We study the optimal liquidation of a large position on Uniswap v2 and Uniswap v3 in discrete time. The instantaneous price impact is derived from the AMM pricing rule. Transient impact is modeled to capture either exponential or approximately power-law decay, together with a permanent component. In the Uniswap v2 setting, we obtain optimal strategies in closed-form under general price dynamics. For Uniswap v3, we consider a two-layer liquidity framework, which naturally extends to multiple layers. We address the problem using dynamic programming under geometric Brownian motion dynamics and approximate the solution numerically using a discretization scheme. We obtain optimal strategies akin to classical ones in the LOB literature, with features specific to Uniswap. In particular, we show how the liquidity profile influences them.
\\


\end{abstract}
\tableofcontents

\newpage

\section{Introduction}

Crypto-assets are primarily traded on two types of venues: Centralized Exchanges (CEXs) and Decentralized Exchanges (DEXs). Whereas CEXs rely on Limit Order Books (LOBs) to match buyers and sellers, DEXs execute transactions on blockchain networks using liquidity pools. Liquidity Providers (LPs) deposit crypto-assets into these pools, while Liquidity Takers (LTs) trade against them. Transaction terms are set by Automated Market Makers (AMMs) through algorithmic pricing rules. Fees are charged on trades and shared between LPs and DEXs.

Uniswap has played a central role in defining the architecture of DEXs. Uniswap v1, launched in 2018, allows users to trade crypto-assets against Ethereum (ETH) and introduced the Constant Product AMM (CPAMM). This mechanism was formalized and generalized in Uniswap v2 \citep{adams2020uniswap} which supports arbitrary crypto-asset pairs. Uniswap v3 \citep{adams2021uniswap} extends the CPAMM by introducing the concept of Concentrated Liquidity AMM (CLAMM). LPs are no longer required to allocate capital across the entire price range; instead, they can specify custom price intervals in which their liquidity is active. Alternative AMMs have emerged to address specific types of crypto-asset pairs, for example Curve \citep{egorov2019stableswap} and Balancer \citep{martinelli2019non}.

The Total Value Locked (TVL) refers to the value of the crypto-assets, generally expressed in dollars, held in liquidity pools. As of September 2025, TVL across all DEXs was $\$25 \text{b}$,\footnote{Data sourced from DefiLlama, available at \url{https://defillama.com/protocols/dexs}.} with an average daily trading volume of $\$12.5 \text{b}$. Uniswap had $\$6 \text{b}$ in TVL and processed an average daily volume of $\$4.5 \text{b}$.

The challenge faced by LTs when executing large trades on LOBs has been extensively studied in the literature. The Almgren-Chriss framework \citep{almgren1999value, almgren2001optimal} pioneers optimal execution modelling by formulating the trade-off between execution costs and the risk of adverse price movements during execution. The model, written in discrete time, combines linear instantaneous and permanent price impacts with a Bachelier dynamic for the fundamental price. The continuous-time counterpart with nonlinear instantaneous impact is developed in \citet{almgren2003optimal}. Several extensions are proposed in \citet{gueant2016financial} and \citet{cartea2015algorithmic}. While the Almgren-Chriss model includes instantaneous and permanent price impacts, it assumes no gradual recovery of the price following a trade. This intermediate regime, known as transient market impact, is accounted for in \citet{obizhaeva2013optimal}; their model has been extended in various directions; see \citet{alfonsi2010optimal, alfonsi2008constrained} and \citet{gatheral2012transient,curato2017optimal}. Notably, \citet{alfonsi2010optimal} generalize the model to arbitrary LOB shape functions. While CPAMMs can be seen as LOBs with virtual shape functions \citep{tran2024order}, the results of \citet{alfonsi2010optimal} do not apply as these shapes depend on the stochastic spot price and the assumption of infinite LOB depth does not hold.

Few works have addressed the optimal execution problem on AMMs. \citet{cartea2025decentralised} develop a continuous-time model in the spirit of Almgren-Chriss, applied to CPAMMs. A first-order approximation is used for the instantaneous price impact, while the permanent impact is assumed to be linear. Different price dynamics are considered, involving both CEX and DEX. However, to the best of our knowledge, the literature has not yet addressed the optimal execution problem on AMMs under transient price impact.

In this paper, we study the optimal execution problem under transient price impact, first on CPAMMs and then CLAMMs. The spot price is modeled as the combination of three components: (i) a fundamental price process; (ii) the cumulative price impact induced by past trades; and (iii) a transient impact model. The instantaneous price impact is derived from the AMM pricing rule. Following \citet{cartea2025decentralised}, we consider a first-order expansion, under which the price impact depends on the square root of the spot price and is therefore stochastic. In the spirit of \citet{obizhaeva2013optimal}, transient price impact is described by a combination of exponential kernels in order to capture either exponential or approximately power-law decay, together with a permanent component. The execution problem is formulated in discrete time and consists in maximizing the expected total cash-flow resulting from liquidation. For CPAMMs, we obtain closed-form solutions under general fundamental price dynamics. For CLAMMs, we consider a two-layer liquidity framework and formulate the problem in a dynamic programming framework under geometric Brownian motion. We then propose a numerical scheme to approximate the solution. Under zero-drift fundamental price dynamics and a first-order approximation, the solution on CPAMMs is independent of the liquidity, whereas on CLAMMs, it adapts to both liquidity levels and the position of the spot price relative to the price threshold where liquidity changes.

The remainder of the paper is organized as follows. Section \ref{section:uniswap_v2} introduces the optimal execution problem on CPAMMs under transient price impact and provides closed-form solutions under general fundamental price dynamics. Anticipating that no closed-form solution is available in the CLAMMs case, the problem is also reformulated and solved within a dynamic programming framework under the geometric Brownian motion assumption. Several scenarios are then considered to illustrate the results. Section \ref{section:uniswap_v3} turns to the CLAMMs case and presents a numerical scheme to approximate the solution. Results are illustrated within a two-layer liquidity framework, which naturally extends to multiple layers.

\section{The optimal scheduling problem on Uniswap v2}\label{section:uniswap_v2}

\subsection{Mathematical framework}

We consider a liquidity pool governed by a Constant Product Automated Market Maker (CPAMM), following the Uniswap v2 design. The pool consists of two tokens (i.e., crypto-assets), denoted by token $a$ (e.g., ETH) and token $b$ (e.g., USDT), with reserves $q^{a}$ and $q^{b}$ respectively. The constant product formula defining the AMM is:
\begin{equation}\label{eq:initial_cpamm}
q^{a} q^{b} = L^{2},
\end{equation}
where $L$ is the liquidity of the pool. The spot price of token $a$ in terms of token $b$ (e.g., the price of ETH in USDT) is given by:
\begin{equation}\label{eq:initial_spot_price}
p = \frac{q^{b}}{q^{a}}.
\end{equation}

Liquidity takers execute trades against the pool. Suppose a trader wants to sell $\delta^{a} > 0$ units of token $a$: the AMM determines the amount $\delta^{b} > 0$ of token $b$ to be received by solving:
\begin{equation}\label{eq:initial_swap}
(q^{a} + \delta^{a}) (q^{b} - \delta^{b}) = L^{2},
\end{equation}
assuming no fees. Fees are discussed in Section \ref{section:fees}. The trade is settled in such a way that the constant product formula \eqref{eq:initial_cpamm} remains satisfied after the swap. Solving \eqref{eq:initial_swap} for $\delta^{b}$ yields:
\begin{equation}\label{eq:initial_formula_delta_b}
\delta^{b} = q^{b} - \frac{L^{2}}{q^{a} + \delta^{a}}.
\end{equation}

From \eqref{eq:initial_cpamm} and \eqref{eq:initial_spot_price}, the reserves can be expressed in terms of liquidity and spot price:
\begin{equation}\label{eq:initial_reserves}
q^{a} = \frac{L}{\sqrt{p}}, \quad q^{b} = L \sqrt{p}.
\end{equation}
Substituting into \eqref{eq:initial_formula_delta_b}, the amount $\delta^{b}$ received by the trader becomes:
\begin{equation}\label{eq:formula_delta_b}
\delta^{b} = \frac{\delta^{a} p}{1 + \frac{\delta^{a} \sqrt{p}}{L}} \approx \delta^{a} p \Big ( 1 - \frac{\delta^{a} \sqrt{p}}{L} \Big ),
\end{equation}
where the approximation corresponds to a first-order Taylor expansion in the dimensionless term $\frac{\delta^{a} \sqrt{p}}{L} = \frac{\delta^{a}}{q^{a}}$. This assumes that the trade size is small relative to the reserve of token $a$.

We now define the execution price of the trade, denoted by $\overline{p}$, as the average exchange rate obtained over the entire trade:
\begin{equation}\label{eq:execution_price}
\overline{p} = \frac{\delta^{b}}{\delta^{a}} = \frac{p}{1 + \frac{\delta^{a} \sqrt{p}}{L}} \approx p \Big ( 1 - \frac{\delta^{a} \sqrt{p}}{L} \Big ),
\end{equation}
where the approximation corresponds to a first-order expansion in $\frac{\delta^{a} \sqrt{p}}{L}$. This represents the effective price paid by the trader per unit of token $a$ over the full trade. In contrast to the spot price $p$, which reflects the marginal price before the trade, the execution price accounts for the cumulative effect of moving along the AMM curve. As emphasized in \citet{cartea2025decentralised}, Equation \eqref{eq:execution_price} is explicitly derived from the AMM pricing rule, in contrast to LOBs, where execution costs are model-dependent. Next, the slippage is defined as the relative deviation of the execution price from the spot price. From the first-order expansion in \eqref{eq:execution_price}, it is approximated by $- \frac{\delta^{a} \sqrt{p}}{L}$. The slope of the slippage with respect to trade size depends on the spot price and therefore becomes stochastic under stochastic spot price dynamics.

The post-swap spot price is given by:
\begin{equation}\label{eq:post_swap_spot_price}
p^{+} = \frac{q^{b} - \delta^{b}}{q^{a} + \delta^{a}} = \frac{p}{\big ( 1 + \frac{\delta^{a} \sqrt{p}}{L} \big )^{2}} \approx p \Big ( 1 - \frac{2 \delta^{a} \sqrt{p}}{L} \Big ),
\end{equation}
where the approximation also corresponds to a first-order expansion in $\frac{\delta^{a} \sqrt{p}}{L}$. This expression reflects the marginal price immediately after the trade and thus incorporates the impact induced by the trade size. The relative price impact is defined as the relative difference between the post-swap spot price and the pre-swap spot price. From the first-order expansion in \eqref{eq:post_swap_spot_price}, it is approximated by $- \frac{2 \delta^{a} \sqrt{p}}{L}$, twice the slippage.

\subsection{Optimal execution strategy}\label{section:optimal_execution strategy_uniswap_v2}

\subsubsection{Problem formulation}\label{section:problem_formulation}

We consider the problem of a trader willing to execute a large sell order of size $\xi > 0$ of token $a$ over a fixed time horizon $T$. The time interval $[0, T]$ is divided into a regular partition: $0 = t_{0} < t_{1} < \cdots < t_{N} = T$, with constant time step $\Delta = \frac{T}{N}$. At each time $t_{n}$, the trader sells an amount $\delta_{n}$ (for the sake of readability, the superscript $a$ is omitted), subject to the volume constraint:
\begin{equation}\label{eq:volume_constraint}
\sum^{N}_{n=0} \delta_{n} = \xi.
\end{equation}
Although the problem is formulated for a large sell order, the large buy order case is analogous.

We assume that the spot price is a combination of three components: (i) a fundamental price process, treated as an exogenous stochastic process capturing market volatility and external factors; (ii) the cumulative price impact induced by previous trades; and (iii) the resilience of the liquidity pool, reflecting the gradual recovery of the spot price following each trade. First, we denote by $(f_{m})^{N}_{m=0}$ the fundamental price process, with $f_{0} > 0$ given as a deterministic initial price. The fundamental price may be interpreted as a reference price, such as the CEX price under the assumption of no friction between CEX and DEX markets. More generally, it corresponds to the spot price from the DEX that would have prevailed in the absence of price impact from the trader’s execution schedule, and thus our model also covers crypto-assets not listed on CEXs. Second, the impact on the spot price at time $t_{m}$ induced by the previous trades $(\delta_{n})^{m-1}_{n=0}$ is derived from the price impact formula \eqref{eq:post_swap_spot_price}. Third, the resilience of the liquidity pool is given by a convex combination of exponential price recovery terms. Hence, we write the spot price at time $t_{m}$, for $m = 0, \ldots, N$, as:
\begin{equation}\label{eq:spot_price_time_m}
p_{m} = f_{m} \Big ( 1 - \sum^{m-1}_{n=0} \sum^{J}_{j=0} \omega_{j} e^{- \rho_{j} (m-n) \Delta} \frac{2 \delta_{n} \sqrt{f_{n}}}{L} \Big ),
\end{equation}
where $(\rho_{j})^{J}_{j=0}$ are resilience parameters and $(\omega_{j})^{J}_{j=0}$ the associated weights, satisfying $\rho_{j} \geq 0$, $\omega_{j} > 0$ for $j = 0, \ldots, J$ and $\sum^{J}_{j=0} \omega_{j} = 1$. A permanent impact component can be accounted for by taking a resilience parameter to zero. Equation \eqref{eq:spot_price_time_m} is constructed by successively applying the relative price impact formula \eqref{eq:post_swap_spot_price} induced by previous trades, combined with a first-order expansion in the dimensionless terms $( \frac{\delta_{n} \sqrt{f_{n}}}{L} )^{m-1}_{n=0}$. Then, we multiply the relative price impacts by the resilience factors $( \sum^{J}_{j=0} \omega_{j} e^{-(m-n) \Delta \rho_{j}} )^{m-1}_{n=0}$. Moreover, the liquidity of the pool is assumed to remain constant throughout the trading horizon, i.e., LPs are passive. This modeling choice is supported by the empirical findings of \citet{cartea2025decentralised}, who report limited LP activity relative to LTs and stable pool liquidity over the relevant execution horizons.

The cash-flow at time $t_{m}$ from selling $\delta_{m}$ units is derived by substituting the spot price from \eqref{eq:spot_price_time_m} into the cash-flow formula \eqref{eq:formula_delta_b} and is also a first-order expansion in the dimensionless terms $( \frac{\delta_{n} \sqrt{f_{n}}}{L} )^{m-1}_{n=0}$:
\begin{equation}\label{eq:cash_tm}
\mathcal{C}_{m} = \delta_{m} f_{m} \Big ( 1 - \sum^{m-1}_{n=0} \sum^{J}_{j=0} \omega_{j} e^{- \rho_{j} (m-n) \Delta} \frac{2 \delta_{n} \sqrt{f_{n}}}{L} - \frac{\delta_{m} \sqrt{f_{m}}}{L} \Big ).
\end{equation}
The execution price that determines the cash-flow at time $t_{m}$ combines two effects both resulting from the trader's actions: the cumulative impact of past trades on the spot price and the slippage induced by the current trade.

We now formulate the execution problem of the trader. The objective is to determine the optimal sequence of trade sizes $\delta = \big ( \delta_{0}, \ldots, \delta_{N} \big )^{\top}$ that maximizes the expected total cash-flow, corresponding to the sum over time of the contributions defined in \eqref{eq:cash_tm}, subject to the volume constraint \eqref{eq:volume_constraint}:
\begin{equation}\label{eq:execution_problem_uniswap_v2}
\begin{aligned}
\delta^{*} = \underset{\delta}{\mathrm{argmax}} \quad & \mathbb{E} \Big [ \sum^{N}_{n=0} \mathcal{C}_{n} \Big ] \\
\textrm{s.t.} \quad & \sum^{N}_{n=0} \delta_{n} = \xi.
\end{aligned}
\end{equation}
The constraint ensures that the full order of size $\xi$ is executed over the time interval $[0, T]$. The trades are not $\textit{a priori}$ required to be non-negative, as depending on the dynamics of the fundamental price, it may be optimal to buy (or oversell relative to $\xi$) and later unwind the excess.

\subsubsection{General solution}\label{section:general_solution}

We now state the general closed-form solution to the execution problem \eqref{eq:execution_problem_uniswap_v2}, formulated using the first-order cash-flow expression \eqref{eq:cash_tm}.
\medskip
\begin{proposition}[General solution]\label{pr:general_solution}
We assume that $(f_{m})_{m=0}^{N}$ is adapted to a filtration $(\mathcal{F}_{m})_{m=0}^{N}$ and satisfies:
\begin{itemize}
\item[] \textnormal{(H1) Integrability.} $\mathbb{E}[ f^{3/2}_{m} ] < \infty$, for $m = 0, \ldots, N$.
\end{itemize}
Then, provided that the matrix $A$ defined below is invertible, the optimal execution schedule is:
\begin{equation}\label{eq:solution_reformulated}
\delta^{*} = \Big ( \xi - \frac{L}{2} \mathbbm{1}^{\top}A^{-1} B \Big ) \frac{A^{-1} \mathbbm{1}}{\mathbbm{1}^{\top} A^{-1} \mathbbm{1}} + \Big ( \frac{L}{2} \mathbbm{1}^{\top}A^{-1} B \Big ) \frac{A^{-1} B}{\mathbbm{1}^{\top}A^{-1} B},
\end{equation}
where $\mathbbm{1} = \big ( 1, \ldots, 1 \big )^{\top}$, the components of the vector $B \in \mathbb{R}^{N+1}$ are defined as:
\begin{equation}\label{eq:vector_B}
B_{m} = \mathbb{E} [ f_{m} ],
\end{equation}
and the entries of the matrix $A \in \mathbb{R}^{(N+1) \times (N+1)}$ are given by:
\begin{equation}\label{eq:matrix_A}
A_{m n} = \left\{
    \begin{array}{ll}
        \sum^{J}_{j=0} \omega_{j} e^{-\rho_{j} (m-n) \Delta} \mathbb{E} [ f_{m} \sqrt{f_{n}} ] & \mbox{if } n \leq m \\
        \sum^{J}_{j=0} \omega_{j} e^{-\rho_{j} (n-m) \Delta} \mathbb{E} [ f_{n} \sqrt{f_{m}} ] & \mbox{if } n > m.
    \end{array}
\right.
\end{equation}
\end{proposition}
Equation \eqref{eq:solution_reformulated} shows that the optimal execution schedule $\delta^{*}$ can be expressed as a linear combination of two distinct profiles: one driven by resilience and expected price impacts ($A^{-1} \mathbbm{1}$), and another that also accounts for the expected price trajectory ($A^{-1} B$). This linear combination reflects the trade-off between minimizing market impact and exploiting expected price dynamics. The respective weights depend on the total order size $\xi$ and the liquidity level $L$.

Proposition \ref{pr:general_solution} does not guarantee that $\delta^{*}$ is the unique maximizer of the execution problem \eqref{eq:execution_problem_uniswap_v2} under general price dynamics; uniqueness is established for the martingale case in Corollary \ref{co:martingale_case}, and for the geometric Brownian motion case in Corollary \ref{co:GBM}.

\medskip
\begin{corollary}[Martingale case]\label{co:martingale_case}
We assume, in addition to \textnormal{(H1)}, that:
\begin{itemize}
\item[] \textnormal{(H2) Positive martingale.} $f_{m} > 0$ a.s. and $\mathbb{E}[ f_{m} \mid \mathcal{F}_{m-1} ] = f_{m-1}$ for $m = 1, \ldots, N$;
\end{itemize}
together with either:
\begin{itemize}
\item[] \textnormal{(H3) Some non-zero resilience.} There exists $j$ such that $\rho_{j} > 0$; \textnormal{or:}
\item[] \textnormal{(H3') Non-degeneracy.} $f_{m} \neq f_{m-1}$ a.s. for $m = 1, \ldots, N$.
\end{itemize}
Then, the optimal solution \eqref{eq:solution_reformulated} is unique and reduces to:
\begin{equation}\label{eq:solution_martingale}
\delta^{*} = \xi \frac{A^{-1} \mathbbm{1}}{\mathbbm{1}^{\top} A^{-1} \mathbbm{1}},
\end{equation}
where the matrix $A$ is defined in \eqref{eq:matrix_A}.
\end{corollary}
In the martingale case, the solution is independent of the liquidity level $L$: although liquidity affects the overall cost of execution, it does not influence the shape of the optimal schedule. The execution strategy is fully determined by the structure of resilience and expected price impacts. Recall that this result is valid only at first order, i.e., when trades are small compared to the pool size.

\medskip
\begin{corollary}[Geometric Brownian motion case]\label{co:GBM}
The fundamental price is assumed to follow a geometric Brownian motion evaluated at discrete times $t_{m}$, for $m = 0, \ldots, N$:
\begin{equation}\label{eq:Geometric_Brownian_motion}
f_{m} = f_{0} e^{(\mu - \frac{\sigma^{2}}{2}) m \Delta + \sigma W_{m}},
\end{equation}
where $\mu$ is the drift, $\sigma > 0$ the volatility and $W_{m}$ denotes the Brownian motion at time $t_{m}$. Under the condition $\mu < \frac{3 \sigma^{2}}{4} + 4 \min_{j} \rho_{j}$, the optimal strategy $\delta^{*}$ is unique; in the martingale case ($\mu = 0$), it is automatically satisfied. Moreover, in the single exponential kernel setting ($J = 0$), this condition is necessary and sufficient for the uniqueness of $\delta^{*}$, and the inverse of matrix $A$ admits a closed-form expression:
\begin{equation}\label{eq:inverse_matrix_A}
A^{-1} = \frac{1}{f_{0} \sqrt{f_{0}}}
\begin{bmatrix}
1 + \frac{a^{2}}{\gamma_{1}} & -\frac{a}{\gamma_{1}} & 0 & \cdots & 0 \\
-\frac{a}{\gamma_{1}} & \frac{1}{\gamma_{1}} + \frac{a^{2}}{\gamma_{2}} & -\frac{a}{\gamma_{2}} & \cdots & 0 \\
0 & -\frac{a}{\gamma_{2}} & \frac{1}{\gamma_{2}} + \frac{a^{2}}{\gamma_{3}} & \cdots & 0 \\
\vdots & \vdots & \vdots & \ddots & -\frac{a}{\gamma_{N}} \\
0 & 0 & 0 & -\frac{a}{\gamma_{N}} & \frac{1}{\gamma_{N}}
\end{bmatrix},
\end{equation}
where $a = e^{-(\rho - \mu) \Delta}$, $b_{n} = e^{\big ( \frac{3 \mu}{2} + \frac{3 \sigma^{2}}{8} \big ) n \Delta}$ and $\gamma_{n} = b_{n} - a^{2} b_{n-1}$ for $n = 1, \ldots, N$.
\end{corollary}
The proofs are provided in Appendix \ref{appendix:proofs}. While the inverse of $A$ can be expressed explicitly in the single exponential kernel case, it must be computed numerically in the general multi-kernel setting.

\subsubsection{Dynamic programming}\label{section:dynamic_programming_closed_loop}

Anticipating that no general solution will be available in the CLAMMs case, we reformulate the execution problem \eqref{eq:execution_problem_uniswap_v2} within a dynamic programming framework. We distinguish between two settings: the closed-loop case, which includes price feedback, and the open-loop case, which does not \citep{kirk2004optimal}. We start with the closed-loop formulation, in which the state variables at time $t_{n}$ are given by the triplet $(x_{n}, (I^{j}_{n})^{J}_{j=0}, f_{n})$, where $x_{n}$ is the remaining inventory, $I^{j}_{n}$ the cumulative price impact induced by the $j$-th resilience factor and $f_{n}$ the fundamental price. The remaining inventory evolves according to:
\begin{equation}\label{eq:dynamic_inventory}
\left\{
    \begin{array}{ll}
        x_{0} = \xi \\
        x_{n+1} = x_{n} - \delta_{n}.
    \end{array}
\right.
\end{equation}
The dynamics of the cumulative price impacts are derived from \eqref{eq:spot_price_time_m} and read:
\begin{equation}\label{eq:dynamic_cumulative_impact_closed_loop}
\left\{
    \begin{array}{ll}
        I^{j}_{0} = 0 \\
        I^{j}_{n+1} = e^{-\rho_{j} \Delta} \Big (I^{j}_{n} + \frac{2 \delta_{n} \sqrt{f_{n}}}{L} \Big ),
    \end{array}
\right.
\end{equation}
for $j = 0, \ldots, J$.

Let $v_{n}(x_{n}, (I^{j}_{n})^{J}_{j=0}, f_{n})$ denote the value function, representing the maximal expected proceeds from liquidating $x_{n}$ units over the periods $t_{n}, \ldots, t_{N}$, given the current cumulative price impacts and fundamental price. From \eqref{eq:execution_problem_uniswap_v2}, the associated Bellman equation reads:
\begin{equation}\label{eq:Bellman_equation_closed_loop_uniswap_v2}
v_{n}(x_{n}, (I^{j}_{n})^{J}_{j=0}, f_{n}) = \sup_{\delta_{n}} \ \delta_{n} f_{n} \big ( 1 - \sum^{J}_{j=0} \omega_{j} I^{j}_{n} - \frac{\delta_{n} \sqrt{f_{n}}}{L} \big ) + \mathbb{E} \Big [ v_{n+1} (x_{n+1}, (I^{j}_{n+1})^{J}_{j=0}, f_{n+1}) | f_{n} \Big ].
\end{equation}
To satisfy the volume constraint \eqref{eq:volume_constraint}, the terminal condition enforces complete liquidation:
\begin{equation}\label{eq:Bellman_equation_terminal_closed_loop_uniswap_v2}
v_{N}(x_{N}, (I^{j}_{N})^{J}_{j=0}, f_{N}) = x_{N} f_{N} \Big ( 1 - \sum^{J}_{j=0} \omega_{j} I^{j}_{N} - \frac{x_{N} \sqrt{f_{N}}}{L} \Big ),
\end{equation}
which corresponds to selling the entire remaining inventory $x_{N}$ at $t_{N}$ in a single trade.

In contrast to Proposition \ref{pr:general_solution}, where the dynamics of the fundamental price remain unspecified, solving the Bellman equation and deriving the value function require specifying them. We assume that the fundamental price follows a geometric Brownian motion \eqref{eq:Geometric_Brownian_motion}. The explicit value function and optimal control in a closed-loop setting are provided in the following proposition.
\medskip
\begin{proposition}[Dynamic programming: closed-loop solution]\label{pr:dynamic_programming_closed_loop}
Under the geometric Brownian motion assumption \eqref{eq:Geometric_Brownian_motion}, the value function reads:
\begin{equation}\label{eq:value_function_closed_loop_uniswap_v2}
v_{n}(x_{n}, (I^{j}_{n})^{J}_{j=0}, f_{n}) = x_{n} f_{n} \Big ( A_{n} + \sum^{J}_{j=0} B^{j}_{n} I^{j}_{n} + C_{n} x_{n} \sqrt{f_{n}} \Big ) + \sqrt{f_{n}} \Big ( D_{n} + \sum^{J}_{j=0} E^{j}_{n} I^{j}_{n} + \sum^{J}_{j_{1}=0} \sum^{J}_{j_{2}=0} F^{j_{1}, j_{2}}_{n} I^{j_{1}}_{n} I^{j_{2}}_{n} \Big ),
\end{equation}
and the optimal control:
\begin{equation}\label{eq:solution_closed_loop_uniswap_v2}
\delta^{*}_{n}(x_{n}, (I^{j}_{n})^{J}_{j=0}, f_{n}) = \frac{1}{2 \phi_{n+1} \sqrt{f_{n}}} \Big [ \theta^{1}_{n+1} + \sum^{J}_{j=0} I^{j}_{n} \theta^{2, j}_{n+1} + x_{n} \sqrt{f_{n}} \theta^{3}_{n+1} \Big ],
\end{equation}
where the coefficients $\phi_{n+1}$, $\theta^{1}_{n+1}$, $\big ( \theta^{2, j_{1}}_{n+1} \big )^{J}_{j_{1}=0}$ and $\theta^{3}_{n+1}$ are defined as:
\begin{equation}\label{eq:phi_closed_loop_uniswap_v2}
\phi_{n+1} = \frac{1}{L} + \frac{2}{L} \sum^{J}_{j=0} B^{j}_{n+1} e^{(\mu-\rho_{j}) \Delta} - C_{n+1} e^{(\frac{3 \mu}{2} + \frac{3 \sigma^{2}}{8}) \Delta} - \frac{4}{L^{2}} \sum^{J}_{j_{1}=0} \sum^{J}_{j_{2}=0} F^{j_{1}, j_{2}}_{n+1} e^{(\frac{\mu}{2} - \rho_{j_{1}} - \rho_{j_{2}} - \frac{\sigma^{2}}{8}) \Delta},
\end{equation}
and,
\begin{equation}
\begin{aligned}\label{eq:thetas_closed_loop_uniswap_v2}
\theta^{1}_{n+1} & = 1 - A_{n+1} e^{\mu \Delta} + \frac{2}{L} \sum^{J}_{j=0} E^{j}_{n+1} e^{(\frac{\mu}{2} - \rho_{j} - \frac{\sigma^{2}}{8}) \Delta}, \\
\theta^{2, j_{1}}_{n+1} & = - \omega_{j_{1}} - B^{j_{1}}_{n+1} e^{(\mu - \rho_{j_{1}}) \Delta} + \frac{4}{L} \sum^{J}_{j_{2}=0} F^{j_{1}, j_{2}}_{n+1}  e^{(\frac{\mu}{2} - \rho_{j_{1}} - \rho_{j_{2}} - \frac{\sigma^{2}}{8}) \Delta}, \\
\theta^{3}_{n+1} & = \frac{2}{L} \sum^{J}_{j=0} B^{j}_{n+1} e^{(\mu - \rho_{j}) \Delta} - 2 C_{n+1} e^{(\frac{3\mu}{2} + \frac{3\sigma^{2}}{8}) \Delta}.
\end{aligned}
\end{equation}
The coefficients $A_{n}, (B^{j}_{n})^{J}_{j=0}, C_{n}, D_{n}, (E_{n})^{J}_{j=0}$ and $(F_{n})^{J}_{j_{1},j_{2}=0}$ are determined recursively as follows:
\begin{equation}\label{eq:parameters_closed_loop_uniswap_v2}
\left\{
    \begin{array}{ll}
    A_{n} = A_{n+1} e^{\mu \Delta} + \frac{1}{2 \phi_{n+1}} \theta^{1}_{n+1} \theta^{3}_{n+1} \\
    B^{j}_{n} = B^{j}_{n+1} e^{(\mu - \rho_{j}) \Delta} + \frac{1}{2 \phi_{n+1}} \theta^{2,j}_{n+1} \theta^{3}_{n+1} \\
    C_{n} = C_{n+1} e^{(\frac{3 \mu}{2} + \frac{3 \sigma^{2}}{8}) \Delta} + \frac{1}{4 \phi_{n+1}} (\theta^{3}_{n+1})^{2} \\
    D_{n} = D_{n+1} e^{(\frac{\mu}{2} - \frac{\sigma^{2}}{8}) \Delta} + \frac{1}{4 \phi_{n+1}} (\theta^{1}_{n+1})^{2} \\
    E^{j}_{n} = E^{j}_{n+1} e^{(\frac{\mu}{2} - \rho_{j} - \frac{\sigma^{2}}{8}) \Delta} + \frac{1}{2 \phi_{n+1}} \theta^{1}_{n+1} \theta^{2,j}_{n+1} \\
    F^{j_{1}, j_{2}}_{n} = F^{j_{1}, j_{2}}_{n+1} e^{(\frac{\mu}{2} - \rho_{j_{1}} - \rho_{j_{2}} - \frac{\sigma^{2}}{8}) \Delta} + \frac{1}{4 \phi_{n+1}} \theta^{2,j_{1}}_{n+1} \theta^{2, j_{2}}_{n+1}, \\
    \end{array}
\right.
\end{equation}
with terminal conditions $A_{N} = 1$, $B^{j}_{N} = -\omega_{j}$, $C_{N} = - \frac{1}{L}$, $D_{N} = E^{j}_{N} = F^{j_{1}, j_{2}}_{N} = 0$.
\end{proposition}
Details on the derivation are provided in Appendix \ref{appendix:proofs}. Whereas the general solution from Proposition \ref{pr:general_solution} is static, the closed-loop solution is dynamic as it adjusts its execution path in response to price feedback.

We now turn to the open-loop formulation of the execution problem \eqref{eq:execution_problem_uniswap_v2}. The fundamental price is removed from the set of state variables and the problem is reformulated in expectation under the geometric Brownian motion assumption \eqref{eq:Geometric_Brownian_motion}. Consequently, the strategy does not incorporate feedback from the evolving fundamental price during execution. We introduce $\tilde{I}^{j}_{n}$, the cumulative mean price impact at time $t_{n}$ induced by the $j$-th resilience factor, for $j = 0, \ldots, J$. The dynamics are:
\begin{equation}\label{eq:dynamic_Ij}
\left\{
    \begin{array}{ll}
        \tilde{I}^{j}_{0} = 0 \\
        \tilde{I}^{j}_{n+1} = e^{- \Delta \rho_{j}} \Big (\tilde{I}^{j}_{n} + \frac{2 \delta_{n} \sqrt{f_{0}}}{L} e^{(\frac{\mu}{2} + \frac{3 \sigma^{2}}{8}) n \Delta} \Big ).
    \end{array}
\right.
\end{equation}
Inventory also evolves according to \eqref{eq:dynamic_inventory}.

Next, we denote the value function by $\tilde{v}$, and the Bellman equation reads:
\begin{equation}\label{eq:Bellman_equation_open_loop_uniswap_v2}
\tilde{v}_{n}(x_{n}, (\tilde{I}^{j}_{n})^{J}_{j=0}) = \sup_{\delta_{n}} \ \delta_{n} f_{0} e^{\mu n \Delta} \big ( 1 - \sum^{J}_{j=0} \omega_{j} \tilde{I}^{j}_{n} - \frac{\delta_{n} \sqrt{f_{0}}}{L} e^{(\frac{\mu}{2} + \frac{3 \sigma^{2}}{8}) n \Delta} \big ) + \tilde{v}_{n+1} (x_{n+1}, (\tilde{I}^{j}_{n+1})^{J}_{j=0}).
\end{equation}
To satisfy the volume constraint \eqref{eq:volume_constraint}, the terminal condition also enforces complete liquidation:
\begin{equation}\label{eq:Bellman_equation_terminal_open_loop_uniswap_v2}
\tilde{v}_{N}(x_{N}, (\tilde{I}^{j}_{N})^{J}_{j=0}) = x_{N} f_{0} e^{\mu N \Delta} \big ( 1 - \sum^{J}_{j=0} \omega_{j} \tilde{I}^{j}_{N} - \frac{x_{N} \sqrt{f_{0}}}{L} e^{(\frac{\mu}{2} + \frac{3 \sigma^{2}}{8}) N \Delta} \big ).
\end{equation}

The explicit value function and optimal control in the open-loop setting are provided in the following proposition.
\medskip
\begin{proposition}[Dynamic programming: open-loop solution]\label{pr:dynamic_programming_open_loop}
Under the geometric Brownian motion assumption \eqref{eq:Geometric_Brownian_motion}, the value function reads:
\begin{equation}\label{eq:value_function_open_loop_uniswap_v2}
\tilde{v}_{n}(x_{n}, (\tilde{I}^{j}_{n})^{J}_{j=0}) = x_{n} f_{0} \Big ( \tilde{A}_{n} + \sum^{J}_{j=0} \tilde{B}^{j}_{n} \tilde{I}^{j}_{n} + \tilde{C}_{n} x_{n} \sqrt{f_{0}} \Big ) + \sqrt{f_{0}} \Big ( \tilde{D}_{n} + \sum^{J}_{j=0} \tilde{E}^{j}_{n} \tilde{I}^{j}_{n} + \sum^{J}_{j_{1}=0} \sum^{J}_{j_{2}=0} \tilde{F}^{j_{1}, j_{2}}_{n} \tilde{I}^{j_{1}}_{n} \tilde{I}^{j_{2}}_{n} \Big ),
\end{equation}
and the optimal control:
\begin{equation}\label{eq:solution_open_loop_uniswap_v2}
\delta^{*}_{n}(x_{n}, (\tilde{I}^{j}_{n})^{J}_{j=0}) = \frac{1}{2 \tilde{\phi}_{n+1} \sqrt{f_{0}}} \Big [ \tilde{\theta}^{1}_{n+1} + \sum^{J}_{j=0} \tilde{I}^{j}_{n} \tilde{\theta}^{2, j}_{n+1} + x_{n} \sqrt{f_{0}} \tilde{\theta}^{3}_{n+1} \Big ],
\end{equation}
where the coefficients $\tilde{\phi}_{n+1}$, $\tilde{\theta}^{1}_{n+1}$, $\big ( \tilde{\theta}^{2, j_{1}}_{n+1} \big )^{J}_{j_{1}=0}$ and $\tilde{\theta}^{3}_{n+1}$ are defined as:
\begin{equation}\label{eq:phi_open_loop_uniswap_v2}
\tilde{\phi}_{n+1} = \frac{1}{L} e^{(\frac{3 \mu}{2} + \frac{3 \sigma^{2}}{8}) n \Delta} + \frac{2}{L} \sum^{J}_{j=0} \tilde{B}^{j}_{n+1} e^{-\rho_{j} \Delta} e^{(\frac{\mu}{2} + \frac{3 \sigma^{2}}{8}) n \Delta} - \tilde{C}_{n+1} - \frac{4}{L^{2}} \sum^{J}_{j_{1}=0} \sum^{J}_{j_{2}=0} \tilde{F}^{j_{1}, j_{2}}_{n+1} e^{-(\rho_{j_{1}} + \rho_{j_{2}}) \Delta} e^{(\mu + \frac{3 \sigma^{2}}{4}) n \Delta},
\end{equation}
and,
\begin{equation}
\begin{aligned}\label{eq:thetas_open_loop_uniswap_v2}
\tilde{\theta}^{1}_{n+1} & = e^{\mu n \Delta} - \tilde{A}_{n+1} + \frac{2}{L} \sum^{J}_{j=0} \tilde{E}^{j}_{n+1} e^{- \rho_{j} \Delta} e^{(\frac{\mu}{2} + \frac{3 \sigma^{2}}{8}) n \Delta}, \\
\tilde{\theta}^{2, j_{1}}_{n+1} & = - \omega_{j_{1}} e^{\mu n \Delta} - \tilde{B}^{j_{1}}_{n+1} e^{- \rho_{j_{1}} \Delta} + \frac{4}{L} \sum^{J}_{j_{2}=0} \tilde{F}^{j_{1}, j_{2}}_{n+1} e^{-(\rho_{j_{1}} + \rho_{j_{2}}) \Delta} e^{(\frac{\mu}{2} + \frac{3 \sigma^{2}}{8}) n \Delta}, \\
\tilde{\theta}^{3}_{n+1} & = \frac{2}{L} \sum^{J}_{j=0} \tilde{B}^{j}_{n+1} e^{- \rho_{j} \Delta} e^{(\frac{\mu}{2} + \frac{3 \sigma^{2}}{8}) n \Delta} - 2 \tilde{C}_{n+1}.
\end{aligned}
\end{equation}
The coefficients $\tilde{A}_{n}, (\tilde{B}^{j}_{n})^{J}_{j=0}, \tilde{C}_{n}, \tilde{D}_{n}, (\tilde{E}_{n})^{J}_{j=0}$ and $(\tilde{F}_{n})^{J}_{j_{1},j_{2}=0}$ are determined recursively as follows:
\begin{equation}\label{eq:parameters_open_loop_uniswap_v2}
\left\{
    \begin{array}{ll}
    \tilde{A}_{n} = \tilde{A}_{n+1} + \frac{1}{2 \tilde{\phi}_{n+1}} \tilde{\theta}^{1}_{n+1} \tilde{\theta}^{3}_{n+1} \\
    \tilde{B}^{j}_{n} = \tilde{B}^{j}_{n+1} e^{- \rho_{j} \Delta} + \frac{1}{2 \tilde{\phi}_{n+1}} \tilde{\theta}^{2,j}_{n+1} \tilde{\theta}^{3}_{n+1} \\
    \tilde{C}_{n} = \tilde{C}_{n+1} + \frac{1}{4 \tilde{\phi}_{n+1}} (\tilde{\theta}^{3}_{n+1})^{2} \\
    \tilde{D}_{n} = \tilde{D}_{n+1} + \frac{1}{4 \tilde{\phi}_{n+1}} (\tilde{\theta}^{1}_{n+1})^{2} \\
    \tilde{E}^{j}_{n} = \tilde{E}^{j}_{n+1} e^{- \rho_{j} \Delta} + \frac{1}{2 \tilde{\phi}_{n+1}} \tilde{\theta}^{1}_{n+1} \tilde{\theta}^{2,j}_{n+1} \\
    \tilde{F}^{j_{1}, j_{2}}_{n} = \tilde{F}^{j_{1}, j_{2}}_{n+1} e^{- (\rho_{j_{1}} + \rho_{j_{2}}) \Delta} + \frac{1}{4 \tilde{\phi}_{n+1}} \tilde{\theta}^{2,j_{1}}_{n+1} \tilde{\theta}^{2, j_{2}}_{n+1}, \\
    \end{array}
\right.
\end{equation}
with terminal conditions $\tilde{A}_{N} = e^{\mu N \Delta}$, $\tilde{B}^{j}_{N} = -\omega_{j} e^{\mu N \Delta}$, $\tilde{C}_{N} = - \frac{1}{L} e^{(\frac{\mu}{2} + \frac{3 \sigma^{2}}{8}) N \Delta}$, $\tilde{D}_{N} = \tilde{E}^{j}_{N} = \tilde{F}^{j_{1}, j_{2}}_{N} = 0$.
\end{proposition}
The open-loop solution is set at time $t_{0}$ and is not adjusted during execution as in the general solution of Proposition \ref{pr:general_solution}. In particular, when the fundamental price follows a geometric Brownian motion, numerical experiments show that the open-loop solution coincides with the solution of Corollary \ref{co:GBM}, which is unsurprising. An analogous result is established by \citet{alfonsi2010optimal} in a distinct framework. Here, the recursive scheme is more complex and prevents a similar derivation.

\subsection{Numerical results}\label{section:numerical_result_v2}

\subsubsection{Illustrations}\label{section:illustrations}

\begin{figure}
    \centering
    \includegraphics[scale = 0.6]{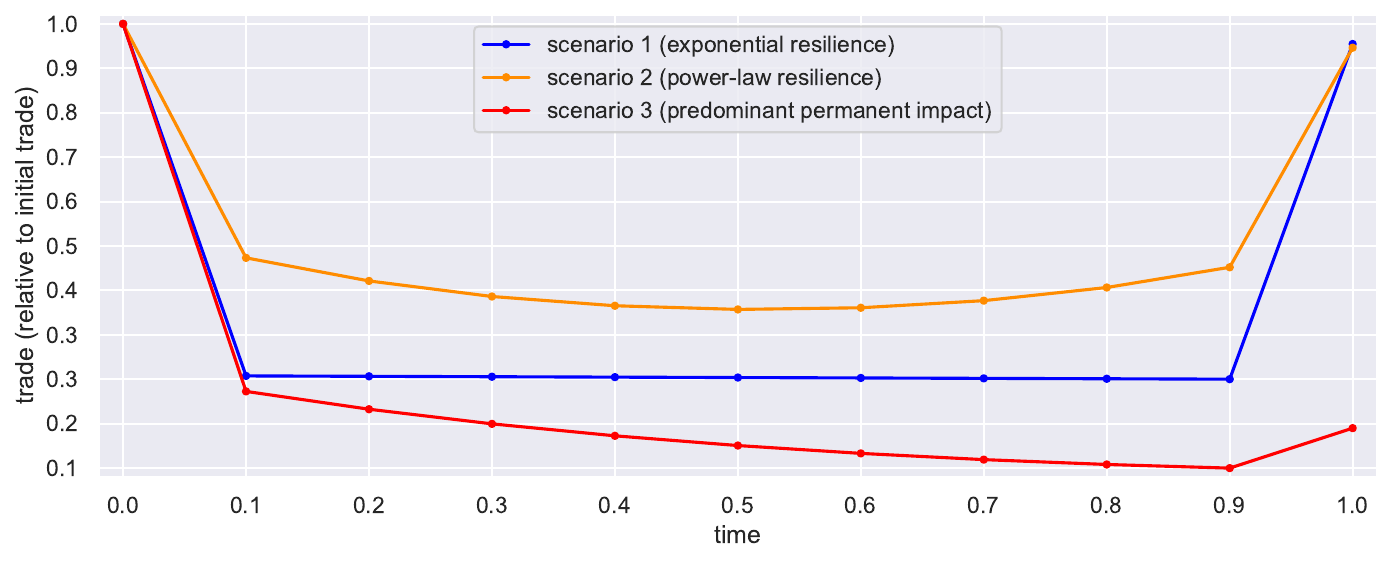}
    \caption{Optimal execution schedule under the three scenarios (relative to the initial trade for the sake of comparison); $T = 1$, $N = 10$, $\xi = 1$, $L = 1000$, $f_{0} = 1$, $\mu = 0$ and $\sigma = 0.3$.}
    \label{fig:toy_examples}
\end{figure}
We illustrate the optimal trading strategy when the fundamental price follows a geometric Brownian motion using the closed-form expression of Corollary \ref{co:GBM} with the following set of parameters: $T = 1$, $N = 10$, $\xi = 1$, $f_{0} = 1$, $L = 1000$, $\mu = 0$ and $\sigma = 0.3$. In the single-kernel setting, the closed-form inverse of $A$ given in \eqref{eq:inverse_matrix_A} is used, whereas in the general multi-kernel case, it is computed numerically. Three scenarios are investigated and Figure \ref{fig:toy_examples} displays the corresponding optimal execution schedules. We compare them to classical ones from the LOB literature. Even though our focus is on optimal execution on CPAMMs, comparisons with the LOB literature remain relevant as CPAMMs can be seen as LOBs with virtual shape functions \citep{tran2024order}.

\paragraph{Scenario 1 (exponential resilience)} We first examine exponential resilience with no permanent impact, setting $J = 0$ and $\rho = 3$. With a single exponential kernel (i.e., $J=0$), the index $j$ in \eqref{eq:spot_price_time_m} is omitted. The resulting optimal strategy exhibits features reminiscent of a bucket-shaped profile, characterized by identical trades at the beginning and end of the trading horizon, and smaller constant sized trades in between. A symmetric bucket-shaped profile was originally observed in \citet{obizhaeva2013optimal} without risk aversion. However, the bucket-shaped profile obtained here is asymmetric, with the initial trade larger than the final one. Appendix \ref{appendix:two_period} establishes this property and studies the sensitivity of the strategy to model parameters, both within a two-period framework. Moreover, the intermediate trades are not equal and exhibit a slight downward slope, a feature also observed in \citet{obizhaeva2013optimal} when risk aversion is included in the objective function. We recover a similar pattern here despite the absence of risk aversion. We interpret this behavior as a consequence of stochastic price impact and slippage, which introduce uncertainty into execution. This asymmetry becomes more pronounced as the volatility parameter increases, reflecting a stronger incentive to front-load execution because of higher expected price impact and slippage of future trades (see Section \ref{section:influences}).

\paragraph{Scenario 2 (power-law resilience)} Next, a power-law-like resilience with no permanent impact is considered. The power-law kernel is approximated by a convex combination of exponential kernels:
\begin{equation}\label{eq:power_law_decay}
(1 + \alpha m \Delta)^{-\beta} \approx \sum^{J}_{j=0} \omega_{j} e^{-\rho_{j} m \Delta},
\end{equation}
where $\alpha \geq 0$ is a scaling parameter and $\beta \geq 0$ denotes the power-law exponent. Given $\alpha$ and $\beta$, the resilience parameters $(\rho_{j})^{J}_{j=0}$ and the associated weights $(\omega_{j})^{J}_{j=0}$ are obtained numerically by fitting the power-law kernel over the interval $[0, T]$. To do so, we use the SLSQP algorithm provided by the SciPy \textit{minimize} function. We set $\alpha = 10$ and $\beta = 0.8$. In this setting, we take $J=1$, which provides an accurate approximation of the power-law kernel. An advanced calibration methodology for larger values of $J$ is presented in \citet{bochud2007optimal}. The optimal execution schedule exhibits an asymmetric U-shaped profile, with trade sizes gradually increasing toward both ends of the trading interval. A symmetric U-shaped profile is presented in \citet[Section 21.2]{bouchaud2018trades} and obtained under power-law decaying propagators without risk aversion whereas an asymmetric U-shaped profile is obtained in \citet{busseti2012calibration} by incorporating risk aversion. As in Scenario $1$, we obtain an asymmetric profile despite the absence of risk aversion. This asymmetry also arises from the stochastic nature of market impact and slippage in our CPAMM framework.

\paragraph{Scenario 3 (predominant permanent impact)} Finally, we consider a scenario under predominant permanent impact and limited exponential resilience, with parameters: $J = 1$, $\omega_{0} = 0.99$, $\rho_{0} = 0$, $\omega_{1} = 0.01$ and $\rho_{1} = 5$. The resulting optimal execution schedule is close to a front-loaded profile, in the spirit of the Almgren-Chriss schedule under a mean-variance objective \citep{almgren2001optimal}. Despite the front-loaded nature of the schedule, the final trade remains significant relative to the previous ones, resulting in a hybrid between a front-loaded and a bucket-shaped profile.

\subsubsection{Sensitivity analysis}\label{section:influences}

\begin{figure}
    \centering
    \begin{subfigure}{\linewidth} 
        \centering
        \includegraphics[scale = 0.6]{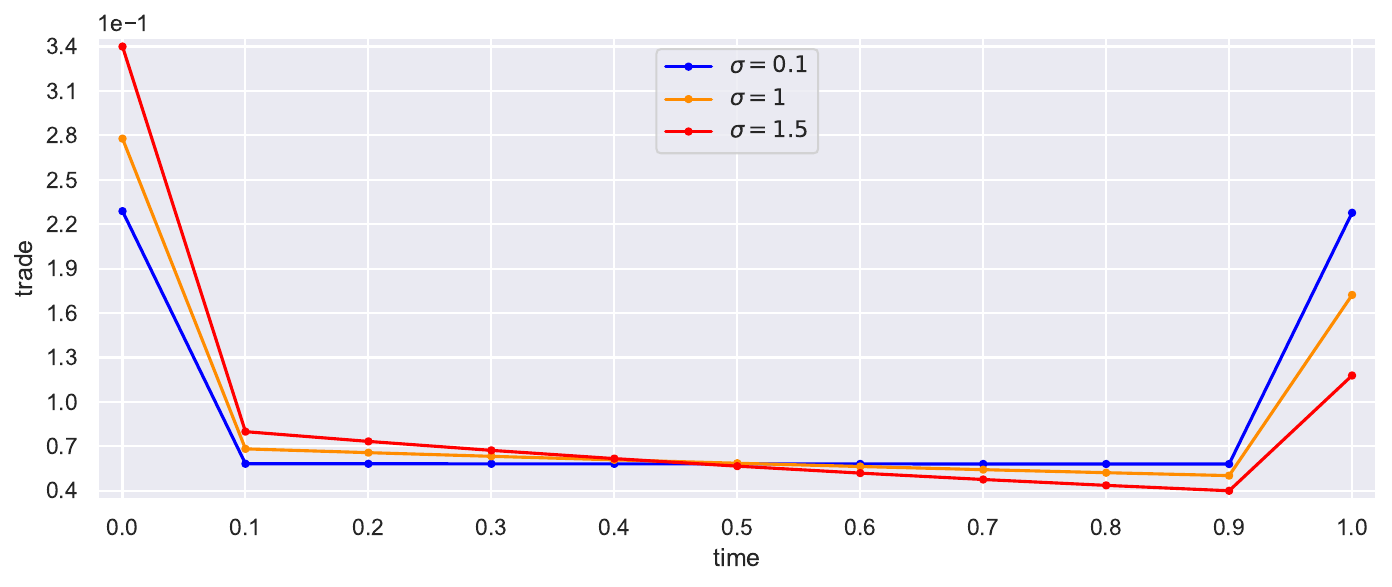}
        \caption{Influence of $\sigma$; $\rho = 3$}
        \label{fig:parameters_influence_0_a}
    \end{subfigure}
    \vfill
    \begin{subfigure}{\linewidth}
        \centering
        \includegraphics[scale = 0.6]{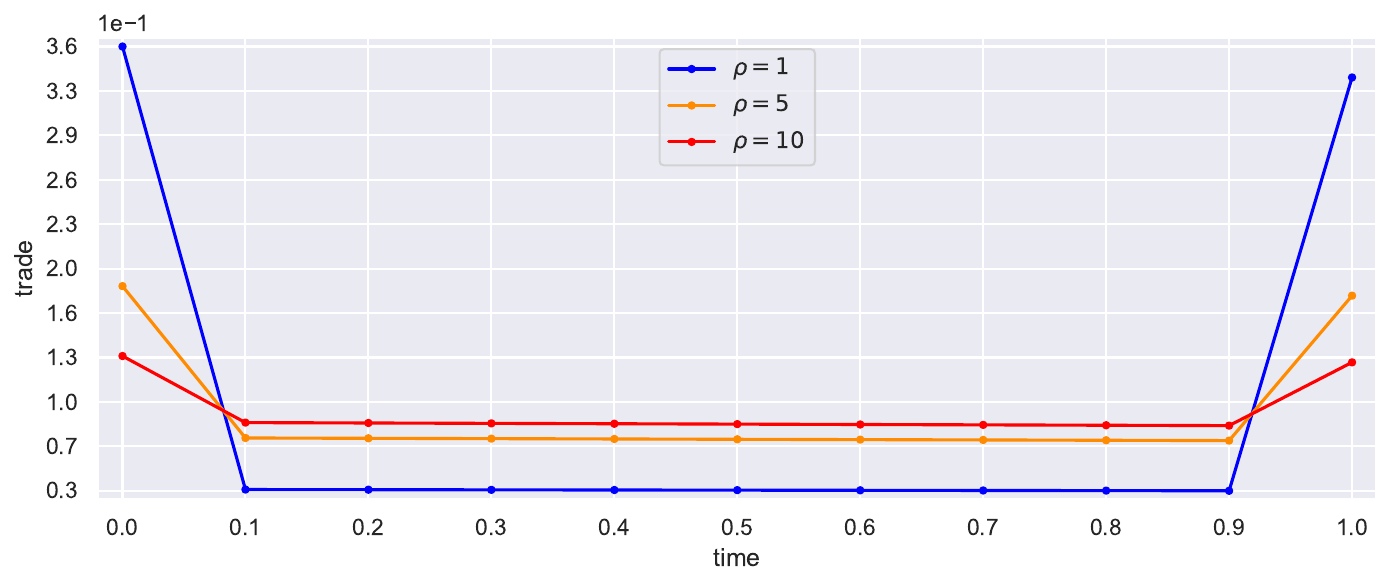}
        \caption{Influence of $\rho$; $\sigma = 0.3$}
        \label{fig:parameters_influence_0_b}
    \end{subfigure}
    \caption{Influence of the parameters on the optimal execution schedule under scenario $1$ (exponential resilience); $T = 1$, $N = 10$, $\xi = 1$, $L = 1000$, $f_{0} = 1$, $\mu = 0$ and $J = 0$.}
    \label{fig:parameters_influence_0}
\end{figure}
\paragraph{Influence of the volatility} In the single-kernel case ($J=0$), corresponding to scenario $1$ (exponential resilience), Figure \ref{fig:parameters_influence_0_a} illustrates the influence of the volatility parameter $\sigma$. As the volatility increases, the execution profile skews toward the beginning of the trading schedule. Indeed, the model allocates larger trade sizes at earlier stages, at the expense of the terminal ones. This behavior reflects the incentive of the trader to front-load execution because of higher expected price impact of future trades.

\begin{figure}
    \centering
    \begin{subfigure}{\linewidth}
        \centering
        \includegraphics[scale = 0.6]{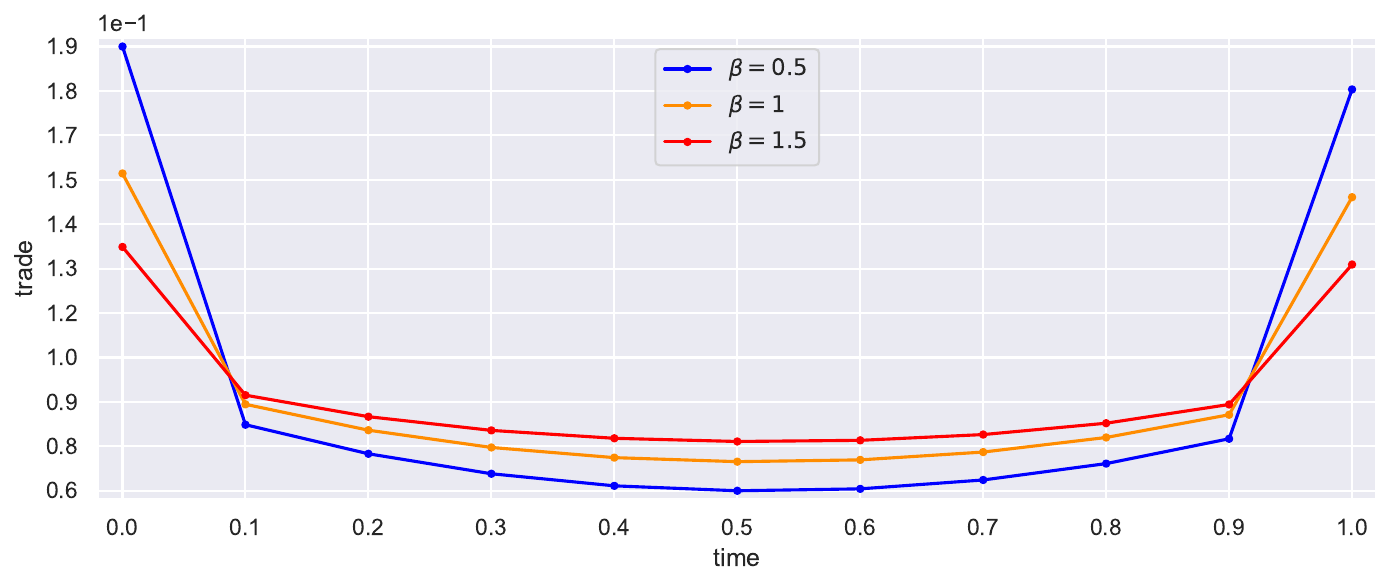}
        \caption{Influence of $\beta$ under scenario $2$ (power-law resilience)}
        \label{fig:parameters_influence_1_a}
    \end{subfigure}
    \vfill
    \begin{subfigure}{\linewidth}
        \centering
        \includegraphics[scale = 0.6]{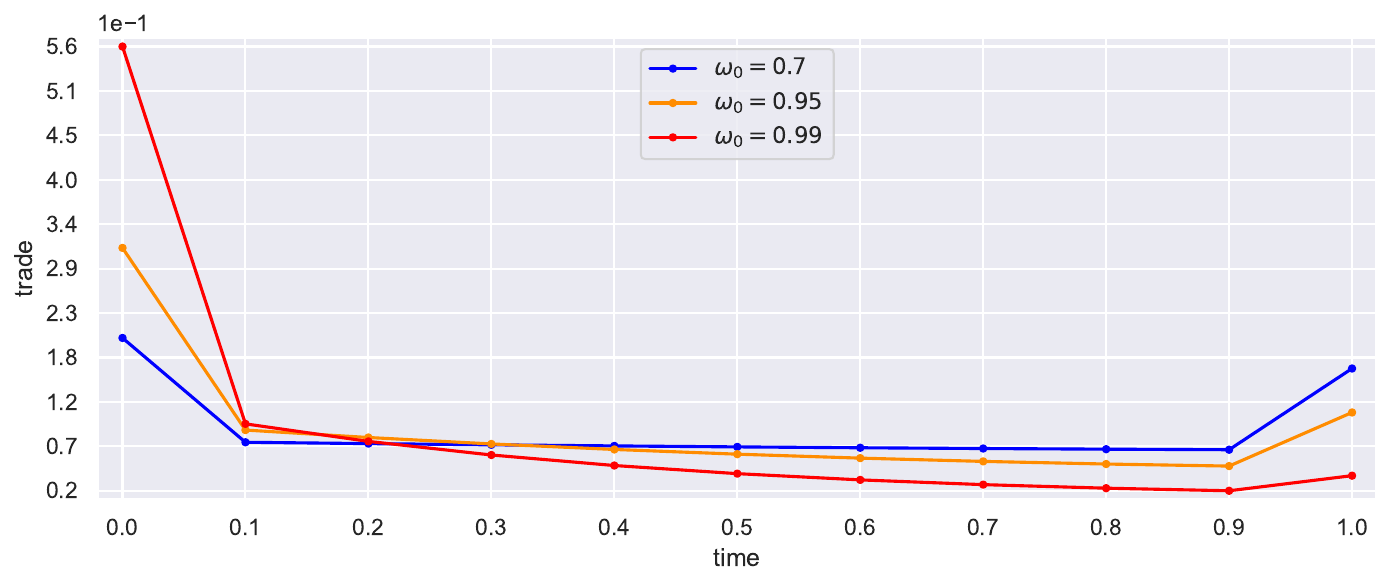}
        \caption{Influence of $\omega_{0}$ under scenario $3$ (predominant permanent impact); $\rho_{0} = 0$, $\rho_{1} = 5$}
        \label{fig:parameters_influence_1_b}
    \end{subfigure}
    \caption{Influence of the parameters on the optimal execution schedule; $T = 1$, $N = 10$, $\xi = 1$, $L = 1000$, $f_{0} = 1$, $\mu = 0$, $\sigma = 0.3$ and $J = 1$.}
    \label{fig:parameters_influence_1}
\end{figure}
\paragraph{Influence of the resilience} Still under scenario $1$ (exponential resilience) in the single-kernel setting ($J = 0$), Figure \ref{fig:parameters_influence_0_b} shows the influence of the resilience parameter $\rho$ on the execution profile. As the resilience increases, the execution profile exhibits an upward parallel shift centered on the intermediate trades. Both the initial and terminal trades decrease in size, while intermediate trades are amplified. This shift enables the strategy to benefit from a stronger mean-reverting effect in prices, thereby reducing overall impact. Then, Figure \ref{fig:parameters_influence_1_a} illustrates the influence of the power-law exponent $\beta$ in the two-kernels setting ($J=1$), corresponding to scenario $2$ (power-law resilience). The resilience parameters are calibrated via the procedure described in Section \ref{section:illustrations}. As $\beta$ increases, the model also increases the size of the intermediate trades. Moreover, a higher $\beta$ concentrates the resilience dynamics around a single dominant factor, thereby flattening the U-shaped profile. Finally, Figure \ref{fig:parameters_influence_1_b} examines the effect of the permanent component of price impact under scenario $3$ (predominant permanent impact). As the permanent impact becomes increasingly predominant, the execution strategy shifts toward a more front-loaded profile. When impact is fully permanent (i.e., $\omega_{0} = 1$), the optimal strategy consists in liquidating the entire inventory at the first time step, as no mean-reverting effect remains to be exploited.

\begin{figure}
    \centering
    \begin{subfigure}{\linewidth} 
        \centering
        \includegraphics[scale = 0.6]{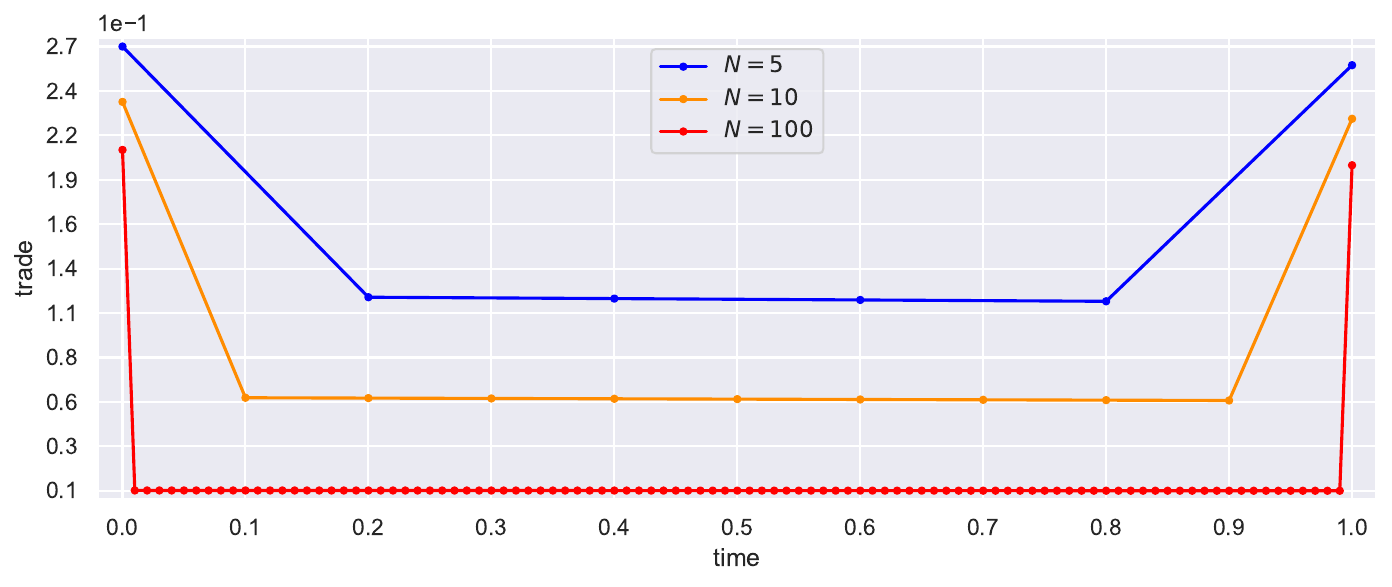}
        \caption{Execution profiles}
        \label{fig:influences_frequency_a}
    \end{subfigure}
    \vfill
    \medskip
    \begin{subfigure}{\linewidth}
        \centering
        \includegraphics[scale = 0.6]{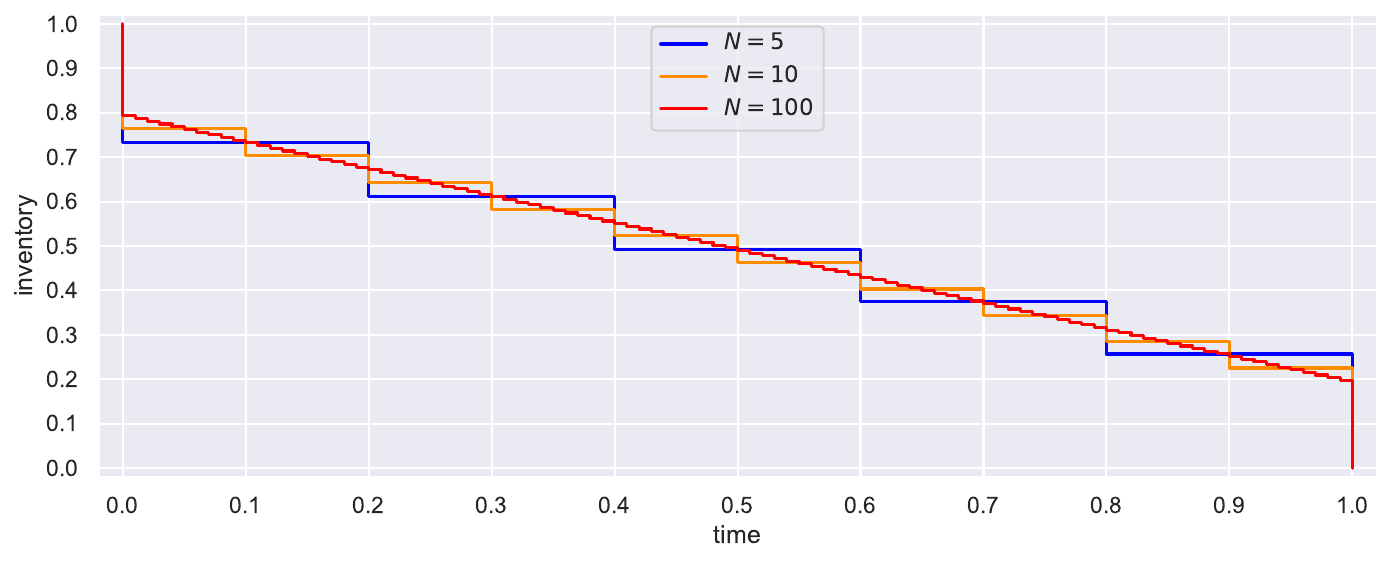}
        \caption{Inventory paths}
        \label{fig:influences_frequency_b}
    \end{subfigure}
    \caption{Influence of the trading frequency on the optimal execution schedule under scenario $1$ (exponential resilience); $T = 1$, $\xi = 1$, $L = 1000$, $f_{0} = 1$, $\mu = 0$, $\sigma = 0.3$, $J = 0$ and $\rho = 3$.}
    \label{fig:influences_frequency}
\end{figure}
\paragraph{Influence of the trading frequency} Figures \ref{fig:influences_frequency_a}, \ref{fig:influences_frequency_b} display the effect of the trading frequency $N$ on both the optimal execution profile and the resulting inventory path in scenario $1$ (exponential resilience). As $N$ increases, all intermediate trades decrease uniformly, and both the initial and final trades also decrease. Nevertheless, the overall structure of the strategy remains unchanged: the inventory paths intersect at all common time steps, indicating that the underlying execution dynamics are preserved.

\begin{figure}
    \centering
    \begin{subfigure}{\linewidth} 
        \centering
        \includegraphics[scale = 0.6]{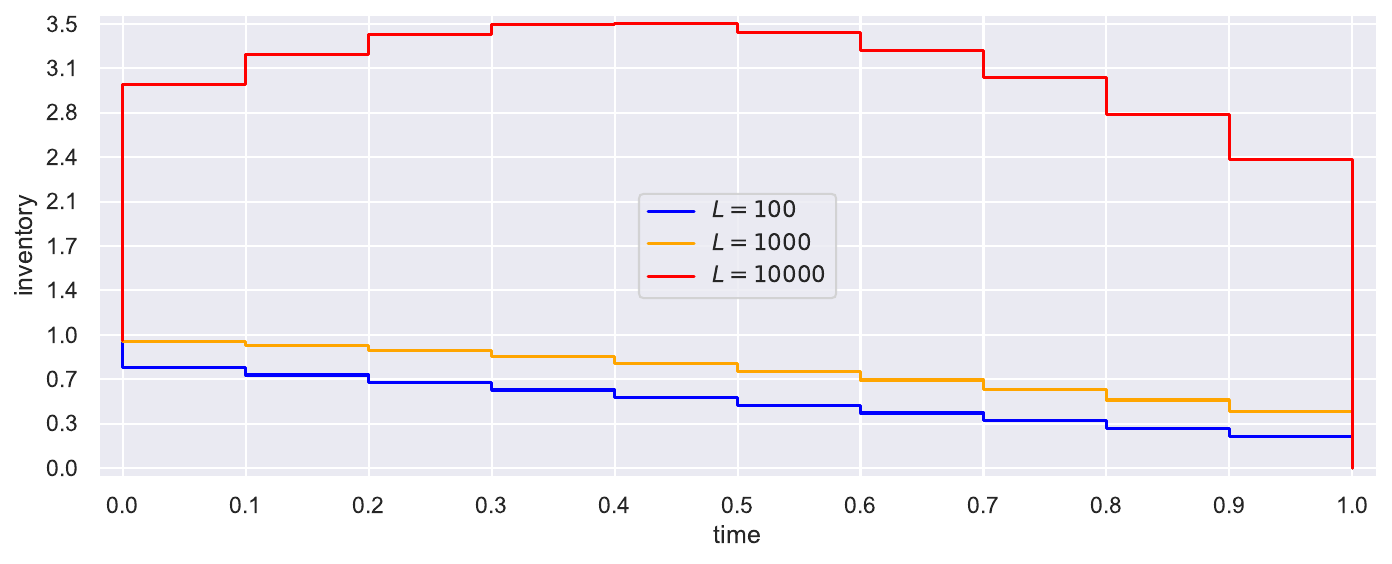}
        \caption{Influence of $L$; $\mu = 0.001$}
        \label{fig:influences_liquidity_a}
    \end{subfigure}
    \vfill
    \medskip
    \begin{subfigure}{\linewidth}
        \centering
        \includegraphics[scale = 0.6]{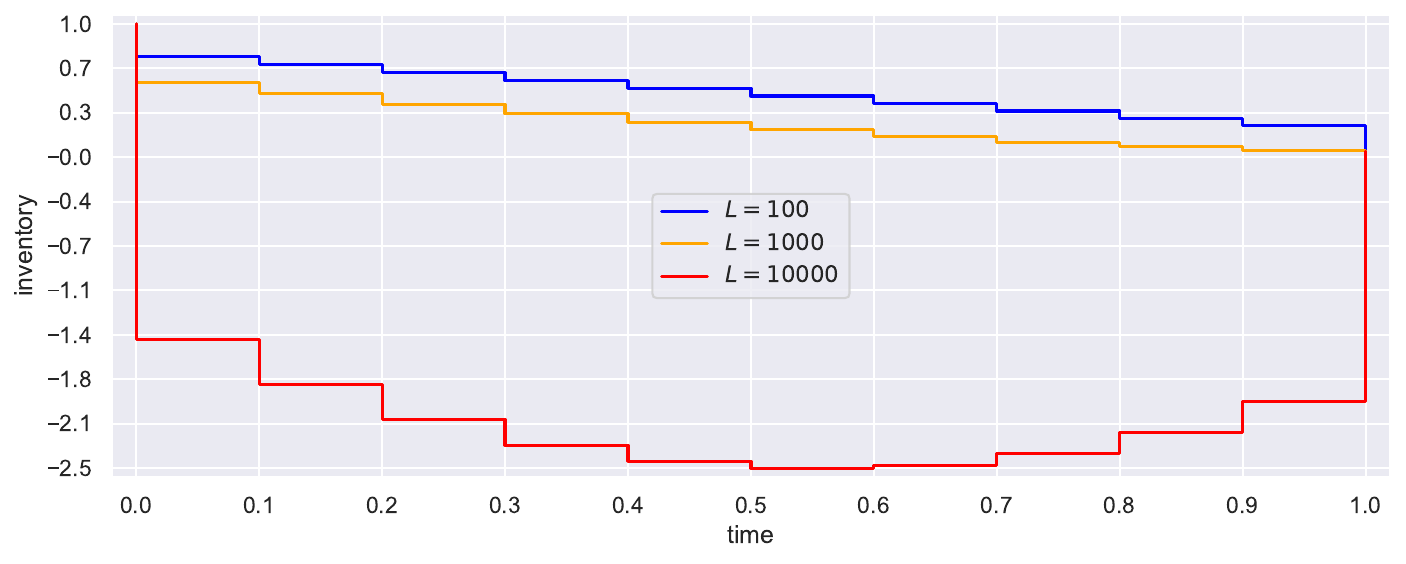}
        \caption{Influence of $L$; $\mu = -0.001$}
        \label{fig:influences_liquidity_b}
    \end{subfigure}
    \caption{Influence of the liquidity with non-zero drift on the optimal execution schedule under scenario $1$ (exponential resilience); $T = 1$, $N = 10$, $\xi = 1$, $f_{0} = 1$, $\sigma = 0.3$, $J = 0$ and $\rho = 3$.}
    \label{fig:influences_liquidity}
\end{figure}
\paragraph{Influence of the liquidity} While liquidity has no impact on the optimal strategy in the martingale setting, it plays a significant role when the drift is non-zero. This effect is illustrated in Figures \ref{fig:influences_liquidity_a}, \ref{fig:influences_liquidity_b}. When the drift is positive, the optimal strategy reduces the size of the early trades and favors liquidation closer to maturity in order to benefit from the expected price increase induced by the positive drift. As the liquidity of the pool increases, the strategy further decreases the size of the early trades, and once liquidity is sufficiently high, the strategy involves buying at the beginning of the period and unwinding the position at maturity. When the drift is negative, the reasoning is reversed and the strategy tends to sell more at the beginning of the period and buy back at maturity in order to benefit from the expected price decrease.

\subsubsection{Open-loop versus closed-loop}\label{section:open_closed}

\begin{table}
\centering
\begin{tabular}{ cccc }
\toprule
& scenario $1$ & scenario $2$ & scenario $3$ \\
\midrule
$\mathcal{E}^{\text{mean}}$ (in basis points) & $3$ & $2$ & $2$ \\
$\mathcal{E}^{\text{max}}$ (in basis points) & $17$ & $9$ & $5$ \\
\bottomrule
\end{tabular}
\caption{Mean and maximum absolute differences between open-loop and closed-loop evaluated along the mean price trajectory optimal execution schedules; $T = 1$, $N = 10$, $\xi = 1$, $L = 1000$, $f_{0} = 1$, $\mu = 0$ and $\sigma = 0.3$.}
\label{tab:open_closed_differences}
\end{table}
First, we compare the optimal strategy obtained in the open-loop setting, as formulated in Proposition \ref{pr:dynamic_programming_open_loop}, with the closed-loop counterpart from Proposition \ref{pr:dynamic_programming_closed_loop} evaluated along the mean fundamental price trajectory. We compute the mean and maximum absolute differences (expressed in basis points) across the three scenarios introduced in Section \ref{section:illustrations}, as summarized in Table \ref{tab:open_closed_differences}. The mean and maximum absolute differences, denoted respectively by $\mathcal{E}^{\text{mean}}$ and $\mathcal{E}^{\text{max}}$, are computed as:
\begin{equation}\label{eq:mean_error}
\mathcal{E}^{\text{mean}} = \frac{1}{N+1} \sum^{N}_{n=0} \Big | \delta^{*}_{n} \big (x_{n}, (I^{j}_{n})^{J}_{j=0}, f_{0} e^{\mu n \Delta} \big ) - \tilde{\delta}^{*}_{n}(x_{n}, (\tilde{I}^{j}_{n})^{J}_{j=0}) \Big |,
\end{equation}
\begin{equation}\label{eq:max_error}
\mathcal{E}^{\text{max}} = \max_{n=0, \ldots, N} \Big | \delta^{*}_{n} \big (x_{n}, (I^{j}_{n})^{J}_{j=0}, f_{0} e^{\mu n \Delta} \big ) - \tilde{\delta}^{*}_{n}(x_{n}, (\tilde{I}^{j}_{n})^{J}_{j=0}) \Big |.
\end{equation}
Both the remaining inventory and the cumulative price impacts are computed forward in time according to their respective dynamics. Across all three scenarios, the open-loop and closed-loop strategies evaluated along the mean fundamental price trajectory yield similar results, with the resulting total cash-flow nearly identical (less than one basis point).

\begin{figure}
    \centering
    \begin{subfigure}{\linewidth} 
        \centering
        \includegraphics[scale = 0.6]{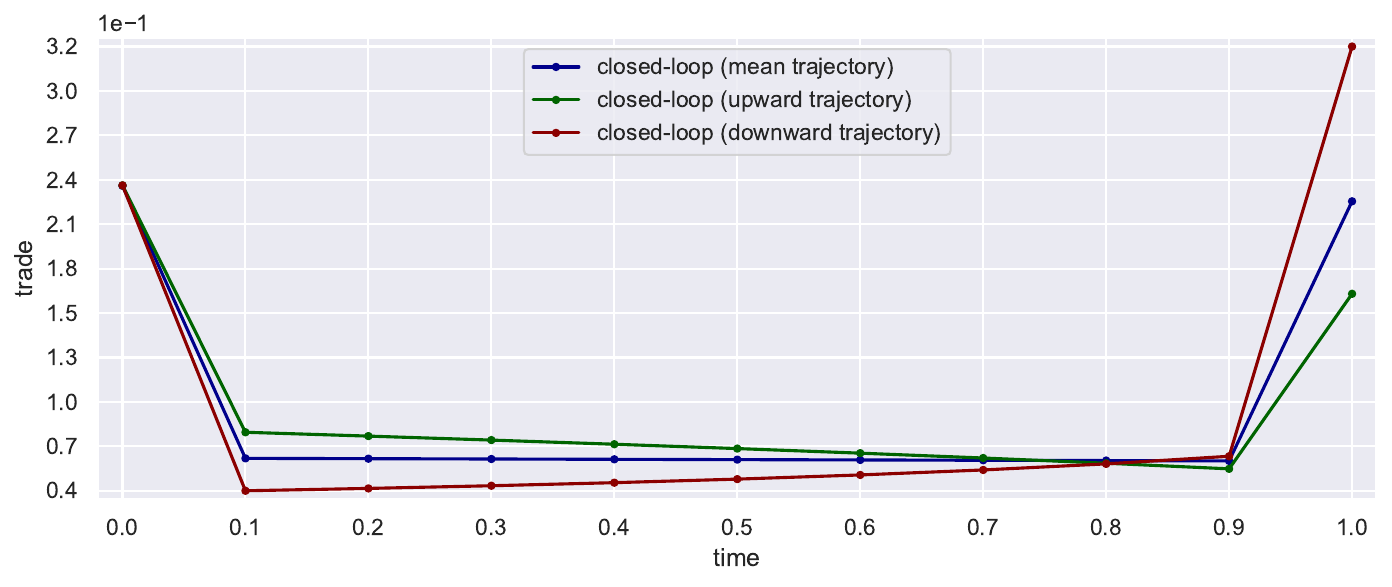}
        \caption{Execution under mean, upward and downward price trajectories}
        \label{fig:open_closed_a}
    \end{subfigure}
    \vfill
    \begin{subfigure}{\linewidth}
        \centering
        \includegraphics[scale = 0.6]{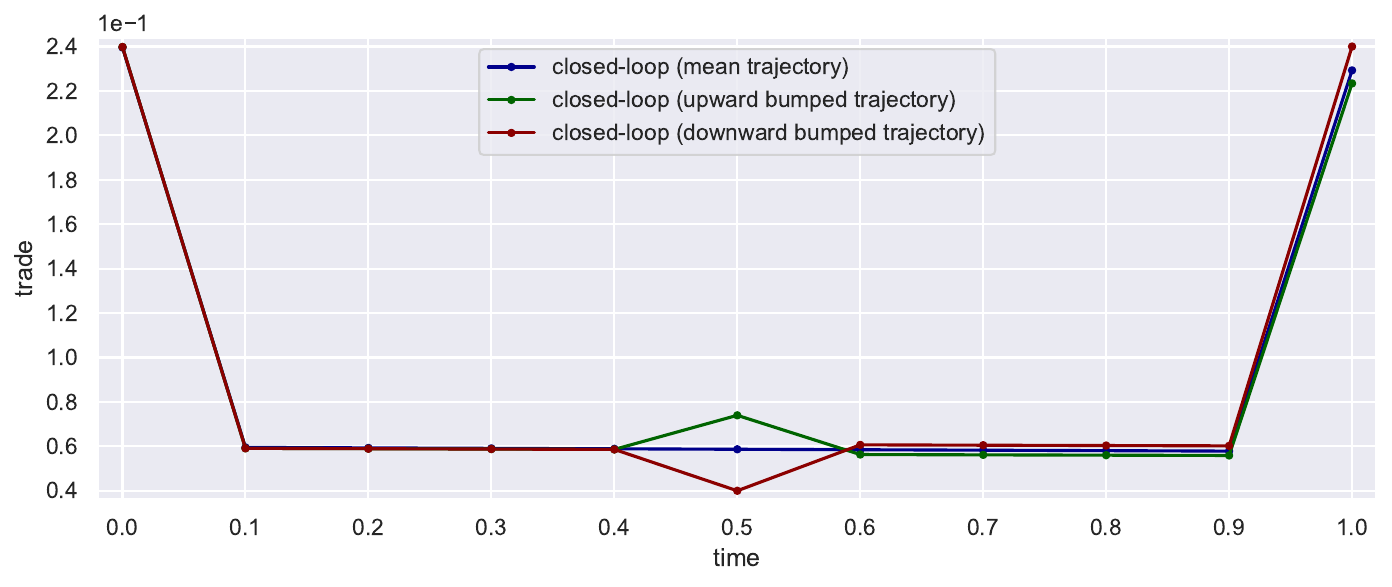}
        \caption{Execution under mean, upward and downward bumped price trajectories}
        \label{fig:open_closed_b}
    \end{subfigure}
    \caption{Closed-loop optimal execution schedules evaluated along mean, upward and downward price trajectories; $T = 1$, $N = 10$, $\xi = 1$, $L = 1000$, $f_{0} = 1$, $\mu = 0$, $\sigma = 0.3$, $J = 0$ and $\rho = 3$.}
    \label{fig:open_closed}
\end{figure}
Second, under scenario $1$ (exponential resilience) from Section \ref{section:illustrations}, we compare the optimal strategy in the closed-loop setting evaluated along the mean fundamental price trajectory with those obtained under the following upward and downward price trajectories:
\begin{equation}\label{eq:upward_downward_trajectories}
f^{\pm}_{m} = f_{0} e^{(\mu - \frac{\sigma^{2}}{2}) m \Delta \pm 3 \sigma m \sqrt{\Delta}}.
\end{equation}
Figure \ref{fig:open_closed_a} presents the resulting execution schedules. At the initial time, the executed trades are identical across trajectories. As prices diverge from the mean trajectory, the strategy adjusts accordingly. In the case of a persistently upward trend, it accelerates liquidation to benefit from rising prices. More specifically, when the observed prices exceed the expected ones, the strategy executes larger trades than it would have under the mean trajectory, seizing the opportunity to sell at more favorable prices. Conversely, when the observed price falls below the mean trajectory, the strategy sells less.

Finally, we analyze the optimal strategy under fundamental price trajectories subject to a persistent upward or downward multiplicative price bump $e^{(\mu - \frac{\sigma^{2}}{2}) \Delta \pm 3 \sigma \sqrt{\Delta}}$, applied just before $t_{n} = 0.5$. The corresponding schedules are displayed in Figure \ref{fig:open_closed_b}. In this case, the strategy reacts immediately at the time of the shock, then returns to the initial profile, shifted to account for changes in remaining inventory.

\subsection{Incorporating fees}\label{section:fees}

In this section, we discuss the integration of a swap fee $\phi$ into the execution framework. The constant product formula \eqref{eq:initial_swap} now reads:
\begin{equation}\label{eq:initial_swap_with_fees}
(q^{a} + (1 - \phi) \delta^{a}) (q^{b} - \delta^{b}) = L^{2}.
\end{equation}
In Uniswap v2, the swap fee is applied to the input amount and is fixed at $\phi = 0.3\%$. After the swap, liquidity is updated to $L^{+} = \sqrt{(q^{a} + \delta^{a}) (q^{b} - \delta^{b})}$ to reflect the fee-adjusted reserves. Equation \eqref{eq:initial_swap_with_fees} is valid only for positive trades (i.e., sell orders), since applying it to negative trades (i.e., buy orders) would imply that fees are received by the trader rather than paid to the pool. In this restricted setting, the optimal scheduling problem \eqref{eq:execution_problem_uniswap_v2} still admits a closed-form solution, with appropriate adjustments to account for fees. Empirically, for the scenarios considered in Section \ref{section:numerical_result_v2}, we observe that a fee level of $\phi = 0.3 \%$ affects cash-flows but has negligible influence on the resulting optimal strategies. Extending the model to handle negative trades would require adjusting the fee structure depending on the sign of the trade, which considerably complicates the modeling framework without giving fundamental new insights. For these reasons, we omit fees in the paper.

\section{From Uniswap v2 to v3}\label{section:uniswap_v3}

\subsection{A two-layer liquidity framework}\label{section:two_layers}

\begin{figure}
    \centering
    \includegraphics[scale = 0.6]{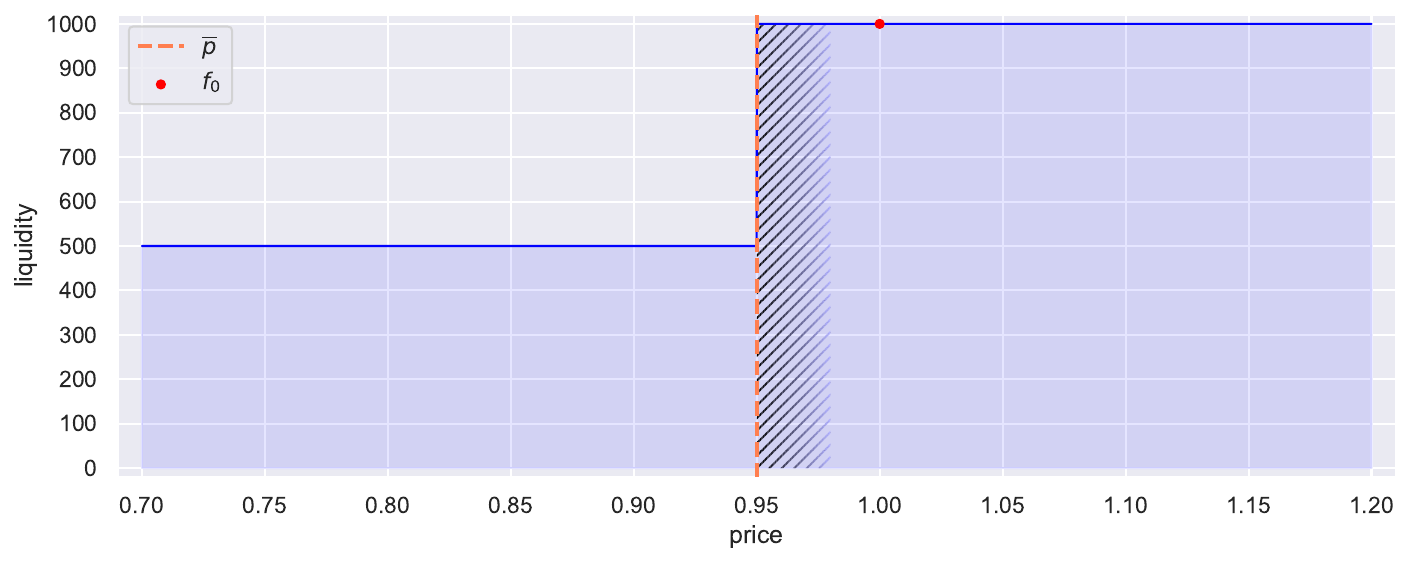}
    \caption{Illustration of the two-layer liquidity profile. For a given trade size, the dashed zone indicates the price interval where the trade may cause the spot price to cross from the high-liquidity layer to the low-liquidity one; $L_{0} = 1000$, $L_{1} = 500$, $f_{0} = 1$ and $\overline{p} = 0.95$.}
    \label{fig:liquidity_v3}
\end{figure}
In Uniswap v3, each LP chooses over which price interval they wish to provide liquidity. As a result, liquidity depends on the spot price in contrast to the constant liquidity of Uniswap v2. Here, we consider a two-layer liquidity framework, which naturally extends to multiple layers: $L_{0}$ when the spot price is above the price threshold $\overline{p}$, and $L_{1} < L_{0}$ when the spot price is below $\overline{p}$. Figure \ref{fig:liquidity_v3} illustrates the liquidity profile as a function of the spot price. We consider the same execution problem \eqref{eq:execution_problem_uniswap_v2}, reformulated within a dynamic programming framework under a closed-loop setting as in Section \ref{section:dynamic_programming_closed_loop}. In the two-layer liquidity framework, both cash-flows and price impacts depend on whether the spot price is above or below the threshold $\overline{p}$. A third case arises when the spot price lies above $\overline{p}$, but the execution causes it to cross the threshold. In this case, both the total cash flows and the aggregate price impact are decomposed into two parts: one corresponding to the execution above the threshold, and the other to the execution below it. The corresponding dynamics of the cumulative price impacts read:
\begin{equation}\label{eq:dynamic_cumulative_impact_closed_loop_v3}
I^{j}_{n+1} = \left\{
    \begin{array}{ll}
    e^{-\rho_{j} \Delta} \big ( I^{j}_{n} + \frac{2 \delta_{n} \sqrt{f_{n}}}{L_{1}} \big ) & \mbox{if } p_{n} \leq \overline{p} \\
    e^{-\rho_{j} \Delta} \big ( I^{j}_{n} + \frac{2 \overline{\delta}_{n} \sqrt{f_{n}}}{L_{0}} + \frac{2 (\delta_{n} - \overline{\delta}_{n}) \sqrt{\overline{p}}}{L_{1}} \big ) & \mbox{if } p_{n} > \overline{p} \mbox{ and } \delta_{n} > \overline{\delta}_{n} \\
    e^{-\rho_{j} \Delta} \big ( I^{j}_{n} + \frac{2 \delta_{n} \sqrt{f_{n}}}{L_{0}} \big ) & \mbox{if } p_{n} > \overline{p} \mbox{ and } \delta_{n} \leq \overline{\delta}_{n},
    \end{array}
\right.
\end{equation}
for $j = 0, \ldots, J$, where $I^{j}_{0} = 0$ and the quantity $\overline{\delta}_{n}$ corresponds to the trade size that brings the post-swap price down to the threshold $\overline{p}$, starting from an initial spot price above the threshold:
\begin{equation}\label{eq:delta_bar_v3}
\overline{\delta}_{n} = \frac{L_{0}}{2 f_{n} \sqrt{f_{n}}} \big ( p_{n} - \overline{p} \big ).
\end{equation}
The dynamics of the inventory remain unchanged from Section \ref{section:dynamic_programming_closed_loop}.

In this framework, the Bellman equation reads:
\begin{equation}\label{eq:Bellman_equation_closed_loop_uniswap_v3}
v_{n}(x_{n}, (I^{j}_{n})^{J}_{j=0}, f_{n}) = \sup_{\delta_{n}} \ \mathcal{C}_{n} + \mathbb{E} \Big [ v_{n+1} (x_{n+1}, (I^{j}_{n+1})^{J}_{j=0}, f_{n+1}) | f_{n} \Big ],
\end{equation}
where the cash-flow at time $t_{n}$ is:
\begin{equation}\label{eq:cash_tn_v3}
\mathcal{C}_{n} = \left\{
            \begin{array}{ll}
                \delta_{n} f_{n} \big ( 1 - \sum^{J}_{j=0} \omega_{j} I^{j}_{n} - \frac{\delta_{n} \sqrt{f_{n}}}{L_{1}} \big ) & \mbox{if } p_{n} \leq \overline{p} \\
                \begin{array}{@{}l}
                \overline{\delta}_{n} f_{n} \big ( 1 - \sum^{J}_{j=0} \omega_{j} I^{j}_{n} - \frac{\overline{\delta}_{n} \sqrt{f_{n}}}{L_{0}} \big ) \\
                \quad \ + \ (\delta_{n} - \overline{\delta}_{n}) \overline{p} \big ( 1 - \sum^{J}_{j=0} \omega_{j} I^{j}_{n} - \frac{(\delta_{n} - \overline{\delta}_{n}) \sqrt{\overline{p}}}{L_{1}} \big )
                \end{array} & \mbox{if } p_{n} > \overline{p} \mbox{ and } \delta_{n} > \overline{\delta}_{n} \\
                \delta_{n} f_{n} \big ( 1 - \sum^{J}_{j=0} \omega_{j} I^{j}_{n} - \frac{\delta_{n} \sqrt{f_{n}}}{L_{0}} \big ) & \mbox{if } p_{n} > \overline{p} \mbox{ and } \delta_{n} \leq \overline{\delta}_{n}.
            \end{array}
        \right.
\end{equation}
To satisfy the volume constraint \eqref{eq:volume_constraint}, the terminal condition of the value function enforces complete liquidation using $\delta_{N} = x_{N}$ in \eqref{eq:cash_tn_v3}.

Because of the discontinuities in the model, closed-form expressions are no longer available. Therefore, in the following section, we present a numerical method to approximate the optimal strategy.

\subsection{Discretization scheme}\label{section:algorithm}

We present a numerical method to approximate the value function and the corresponding optimal control under the two-layer liquidity framework introduced in Section \ref{section:two_layers}. We also assume that the fundamental price follows a geometric Brownian motion, as described in Section \ref{section:dynamic_programming_closed_loop}. We adopt a discretization scheme, starting with the price domain. The price grid is defined for $k_{f} \in \{ 0, \ldots, K_{f} \}$ as:
\begin{equation}\label{eq:grid_p}
f^{(k_{f})} = e^{Y^{(k_{f})}}, \quad Y^{(k_{f})} = m_{T} + z \sigma_{T} \Big (\frac{2 k_{f} - K_{f}}{K_{f}} \Big ),
\end{equation}
where $z > 0$ controls the grid width, while $m_{T} = \log(f_{0}) + (\mu - \frac{\sigma^{2}}{2}) T$ and $\sigma_{T} = \sigma \sqrt{T}$ correspond to the mean and standard deviation of the log-price at maturity. An alternative would be to use adaptive grids based on the mean and standard deviation of the log-price over time. We tested both fixed and adaptive grids: fixed grids yield more stable numerical results.

Next, we discretize the inventory and cumulative price impact spaces. For $k_{x} \in \{ 0, \ldots, K_{x} \}$, the inventory grid is defined by:
\begin{equation}\label{eq:grid_x}
x^{(k_{x})} = \frac{k_{x} \xi}{K_{x}},
\end{equation}
and for $k^{j}_{i} \in \{ 0, \ldots, K_{I} \}$, the grid associated with the cumulative impact induced by the $j$-th resilience factor is given by:
\begin{equation}\label{eq:grid_i}
I^{(k^{j}_{i})} = e^{-\rho_{j} \Delta} \frac{2 k^{j}_{i} \xi \sqrt{f^{(K_{f})}}}{L_{1} K_{I}}.
\end{equation}

To approximate the conditional expectation in the Bellman equation, we introduce log-price midpoints between each $Y^{(k_{f})}$:
\begin{equation}\label{eq:grid_y}
y^{(k^{\pm}_{f})} = \frac{1}{2} \big ( Y^{(k_{f})} + Y^{(k_{f} \pm 1)} \big ),
\end{equation}
with boundary values $y^{(0^{-})} = -\infty$ and $y^{(K^{+}_{f})} = +\infty$ to encompass the entire price domain. Given a current price $f^{(k)}$, the conditional expectation of the value function at the next time step is approximated by:
\begin{equation}\label{eq:cond_expectation}
\mathbb{E} \big [ v_{n+1} (x_{n+1}, (I^{j}_{n+1})^{J}_{j=0}, f_{n+1}) | f_{n} = f^{(k)} \big ] \approx \sum^{K_{f}}_{q=0} \omega_{k \rightarrow q} v_{n+1} (x_{n+1}, (I^{j}_{n+1})^{J}_{j=0}, f^{(q)}),
\end{equation}
where the transition probability weights $\omega_{k \rightarrow q}$ are computed as:
\begin{equation}
\omega_{k \rightarrow q} = \Phi(\frac{y^{(q+)} - \mu_{k}}{s}) - \Phi(\frac{y^{(q-)} - \mu_{k}}{s}),
\end{equation}
with $\mu_{k} = \log(f^{(k)}) + (\mu - \frac{\sigma^{2}}{2}) \Delta$, $s = \sigma \sqrt{\Delta}$ and $\Phi$ denotes the standard normal cumulative distribution function.

The algorithm proceeds backward in time. The terminal condition of the value function is given by \eqref{eq:cash_tn_v3} using $\delta_{N} = x_{N}$. Then, starting from this terminal condition and applying the Bellman recursion, we estimate the value function and the corresponding optimal control at each grid point of the fundamental price, inventory and cumulative impacts, using the approximation of the conditional expectation \eqref{eq:cond_expectation}. Moreover, for a given inventory level indexed by $k_{x}$, the supremum in the Bellman equation \eqref{eq:Bellman_equation_closed_loop_uniswap_v3} is evaluated over a discrete set of trade sizes drawn from the inventory grid: $x^{(k)}$ for $k \in { 0, \ldots, k_{x} }$. This restriction enforces non-negativity of trades, thereby introducing no-buy constraints into the optimization. This design choice is motivated by computational efficiency and by the empirical observation that none of the optimal profiles in the martingale case involve buy orders (see Section \ref{section:numerical_result_v2}). We therefore restrict the use of the algorithm to the martingale case ($\mu = 0$). Taking $\mu \neq 0$ and allowing for buy orders would simply require extending the grid over which the supremum is computed. Appendix \ref{appendix:algorithm} provides a pseudocode of the backward algorithm. In Appendix \ref{appendix:checks}, we present consistency checks within the Uniswap v2 framework to empirically validate our numerical method. Specifically, we compare the numerical results in the particular case where $L_{0} = L_{1}$ with the closed-form solution derived in Section \ref{section:dynamic_programming_closed_loop}, both evaluated along the mean price trajectory. Results show that the discretization scheme yields accurate estimates of the optimal strategy in the Uniswap v2 framework. In Appendix \ref{appendix:exact_dynamic_programming}, we further compare the numerical solution obtained under the first-order cash-flow and price impact formulas with that obtained under their exact counterparts. The resulting deviation is negligible, confirming the validity of the first-order approximation as long as trades are small relative to the reserve of the token being sold, which is the case for the set of parameters considered.

\subsection{Numerical results}\label{section:numerical_results_v3}

Let us analyze the impact of the threshold price $\overline{p}$ relative to the initial price $f_{0}$ on the optimal execution strategy evaluated along the mean price trajectory. The schedules are computed using the discretization scheme described in Section \ref{section:algorithm}, and correspond to the solution to the dynamic programming problem under the two-layer liquidity profile introduced in Section \ref{section:two_layers}. We adopt a grid resolution of $K_{f} = 500$, $K_{x} = 250$ and $K_{I} = 500$. In Appendix \ref{appendix:checks}, we vary the grid sizes to assess the stability of the numerical scheme. The threshold level is expressed in terms of a spread $\overline{s}$, defined by:
\begin{equation}\label{eq:s_bar}
\overline{p} = f_{0} + \overline{s}.
\end{equation}
We distinguish three regimes depending on the position of the threshold $\overline{p}$ relative to the initial price $f_{0}$.
\begin{figure}
    \centering
    \begin{subfigure}{\linewidth} 
        \centering
        \includegraphics[scale = 0.6]{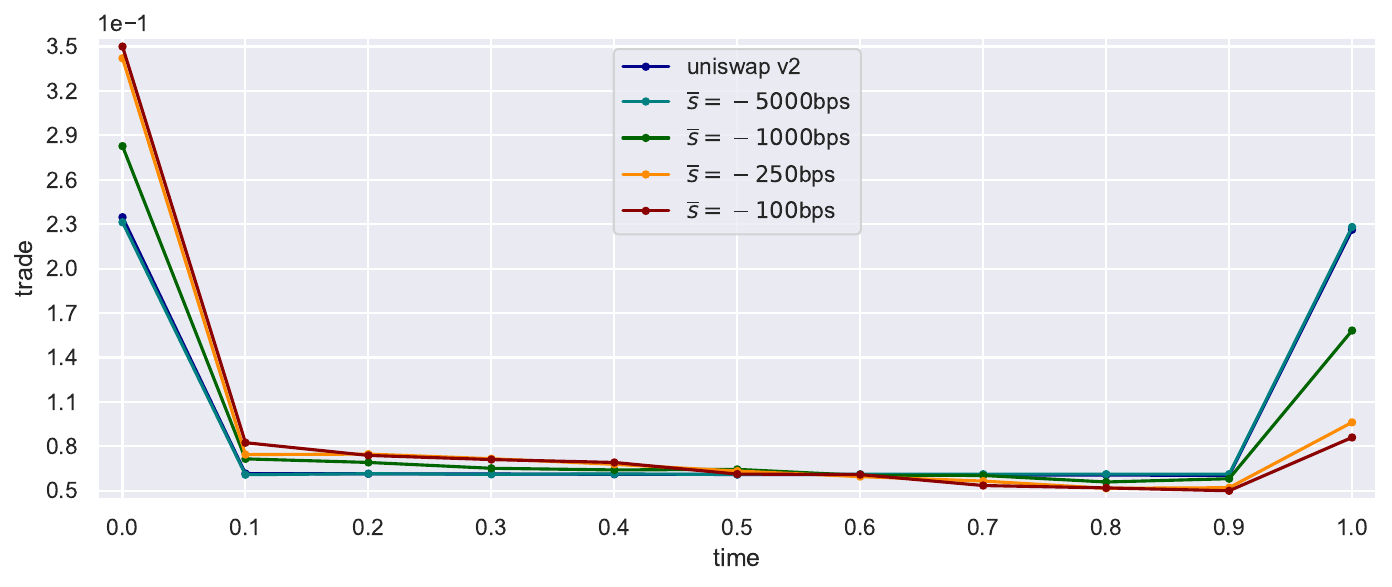}
        \caption{Low price threshold regime}
        \label{fig:optimal_trades_v3_a}
    \end{subfigure}
    \vfill
    \medskip
    \begin{subfigure}{\linewidth}
        \centering
        \includegraphics[scale = 0.6]{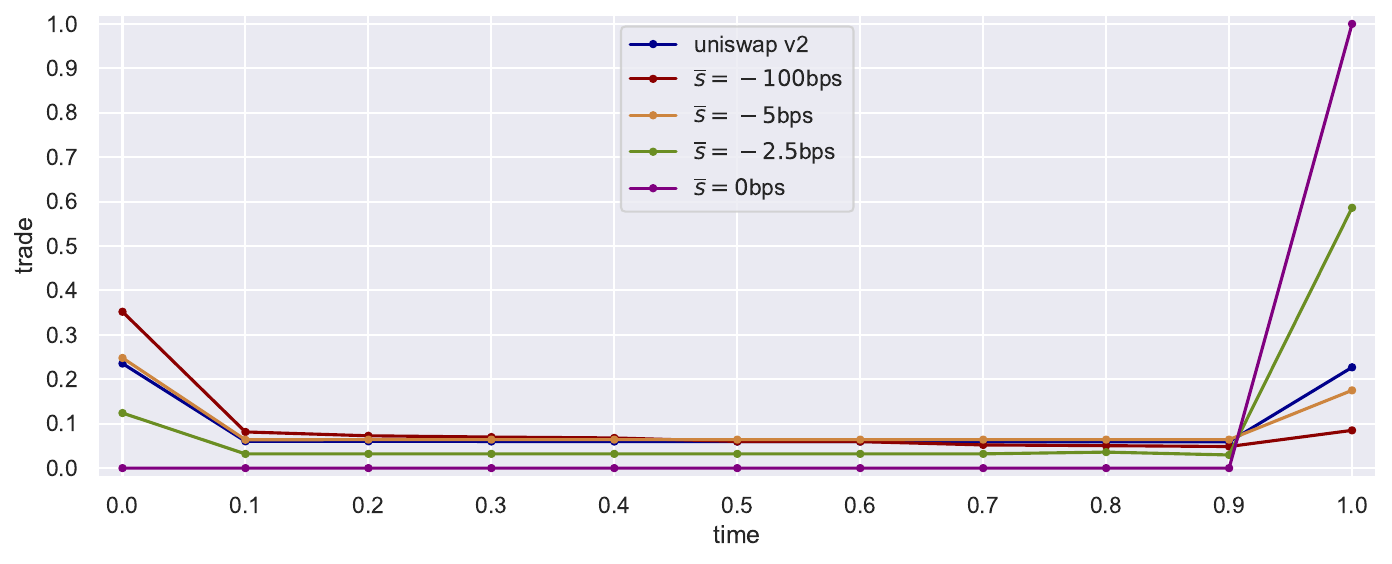}
        \caption{Regime with a price threshold close to the initial spot price}
        \label{fig:optimal_trades_v3_b}
    \end{subfigure}
    \vfill
    \medskip
    \begin{subfigure}{\linewidth}
        \centering
        \includegraphics[scale = 0.6]{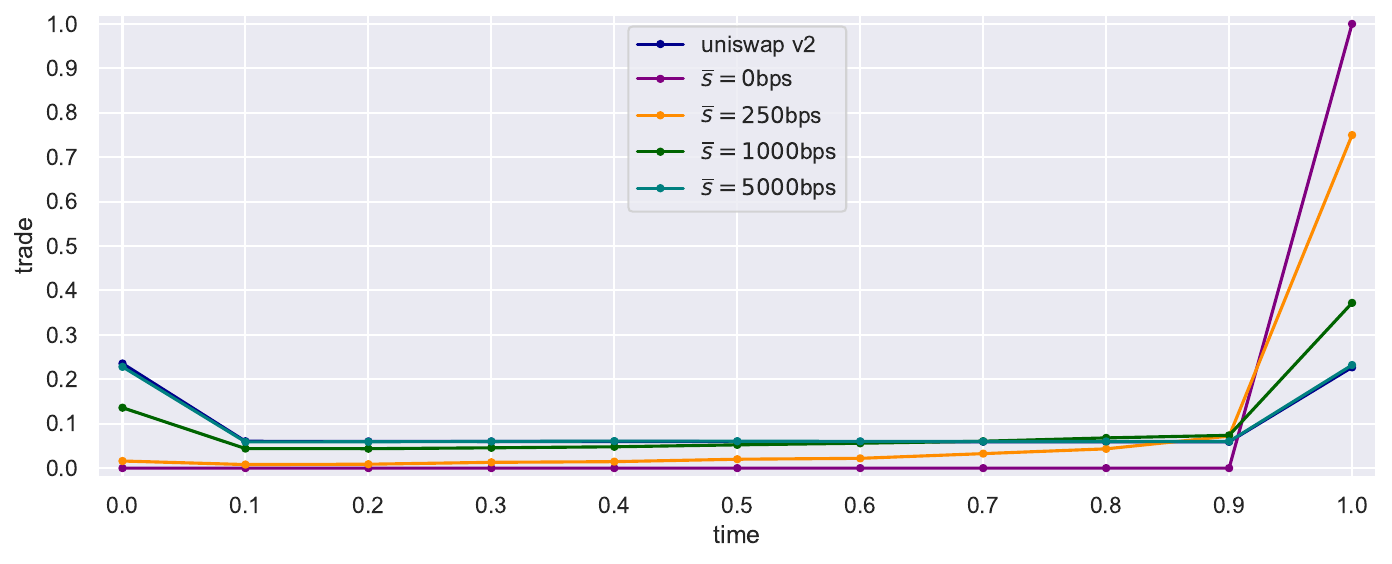}
        \caption{High price threshold regime}
        \label{fig:optimal_trades_v3_c}
    \end{subfigure}
    \caption{Influence of the spread $\overline{s}$ on the optimal execution schedule; $T = 1$, $N = 10$, $\xi = 1$, $L_{0} = 1000$, $L_{1} = 500$, $f_{0} = 1$, $\mu = 0$, $\sigma = 0.3$, $J = 0$, $\rho = 3$, $K_{f} = 500$, $K_{x} = 250$ and $K_{I} = 500$.}
    \label{fig:optimal_trades_v3}
\end{figure}

\paragraph{Low price threshold regime} In Figure \ref{fig:optimal_trades_v3_a}, we present the execution strategies for spreads ranging from $-5000$bps to $-100$bps. When the threshold price lies far below the initial price ($\overline{s} = -5000$bps), the optimal execution strategy coincides with the strategy obtained in the Uniswap v2 framework. Indeed, the probability that the price reaches the low-liquidity zone before maturity is very low. Consequently, the model disregards the discontinuity in liquidity and proceeds as if the pool exhibited constant liquidity. Then, as the threshold gradually approaches $f_{0}$, the execution strategy progressively front-loads the trades. Indeed, we observe a skew in the execution profile, with more volume executed in the early steps of the horizon. This adjustment reflects the increasing probability, driven by price volatility, that the threshold will be crossed before maturity. In anticipation of the associated cost increase in the low-liquidity regime, the model accelerates execution to mitigate potential higher slippage and price impact.

\begin{figure}
    \centering
    \includegraphics[scale = 0.6]{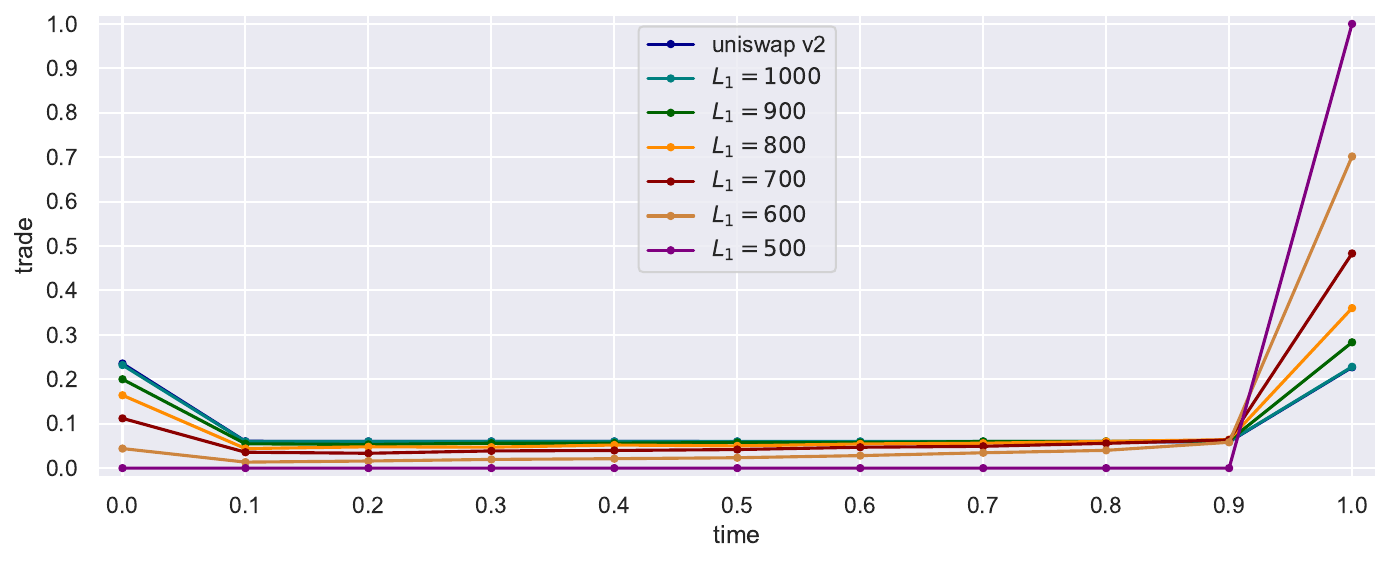}
    \caption{Influence of $L_{1}$ on the optimal execution schedule; $T = 1$, $N = 10$, $\xi = 1$, $L_{0} = 1000$, $f_{0} = 1$, $\overline{s} = 0$, $\mu = 0$, $\sigma = 0.3$, $J = 0$, $\rho = 3$, $K_{f} = 500$, $K_{x} = 250$ and $K_{I} = 500$.}
    \label{fig:impact_L1}
\end{figure}
\paragraph{Regime with a price threshold close to the initial spot price} The execution strategies for spreads ranging from $-100$bps to $0$bps are reported in Figure \ref{fig:optimal_trades_v3_b}. When the threshold level is equal to the initial price ($\overline{s} = 0$), the optimal execution strategy consists of full liquidation at maturity. As previously mentioned, the strategy is evaluated along the mean price trajectory, which, in this case, remains at the threshold throughout the trading horizon. Consequently, the trader holds his position, waiting for a favorable price movement to benefit from lower costs in the higher-liquidity zone. Figure \ref{fig:impact_L1} illustrates the influence of the lower-layer liquidity parameter $L_{1}$ on the optimal strategy. As $L_{1}$ approaches $L_{0}$, the optimal strategy gradually transitions from full liquidation at maturity to the Uniswap v2 optimal execution schedule when $L_{1}$ reaches $L_{0}$.

\paragraph{High price threshold regime} Figure \ref{fig:optimal_trades_v3_c} illustrates the execution profiles for spreads ranging from $0$bps to $5000$bps. Strategies for spreads under $100$bps are omitted as they yield similar results. When the threshold price is set above the initial price, the execution strategy shifts again toward earlier liquidation. The probability of entering the low-liquidity region becomes lower, and the model gradually converges toward the same constant-liquidity execution profile. Indeed, when the fundamental price is a martingale, the optimal strategy under constant liquidity is independent of the liquidity level (see Section \ref{section:optimal_execution strategy_uniswap_v2}).

\section{Discussion and conclusion}

In this paper, we study the optimal execution problem on AMMs, aiming to maximize the sum of expected cash-flows under transient price impact. The study covers both CPAMMs (Uniswap v2) and CLAMMs (Uniswap v3). We derive closed-form solutions under general fundamental price dynamics for CPAMMs, and analyze the resulting strategies when the fundamental price follows a geometric Brownian motion. We also solve the reformulated problem within a dynamic programming framework to allow for price feedback during execution. For CLAMMs, closed-form solutions are no longer available and the optimal strategy is then approximated numerically using a discretization scheme. The analysis is developed within a two-layer liquidity framework, which naturally extends to multiple layers. Numerical results are obtained using synthetic data. The application to real market data, including the calibration of resilience kernels, will be conducted in a companion paper.

In the CPAMMs case, exponential resilience leads to an asymmetric bucket-shaped execution profile, while power-law-like resilience results in an asymmetric U-shaped profile. In the literature, such asymmetries typically stem from risk aversion. In our setting, they arise from the endogenous and stochastic nature of market impact and slippage. Incorporating a mean-variance objective function following the approach of \citet{almgren2001optimal} would enrich the model from a risk-management perspective but would pose significant mathematical challenges which we leave for future investigation.

The optimal execution strategy on CPAMMs, based on a first-order approximation, is independent of the liquidity level of the pool in the martingale setting. In contrast, under the two-layer liquidity profile of CLAMMs, the strategy depends on both the liquidity levels and the position of the spot price relative to the price threshold at which the liquidity changes. When liquidity is more favorable, the strategy tends to front-load execution to reduce exposure to the risk of falling into the low-liquidity region following an adverse price movement, where execution costs are higher. Conversely, when the liquidity level is less favorable, the strategy postpones execution, waiting for a favorable price movement to benefit from lower costs in the higher-liquidity zone. We believe that this work may help practitioners in the DeFi ecosystem better understand execution mechanisms and reduce associated costs.

Finally, we assume that the liquidity of the pool is constant throughout the trading horizon, i.e., LPs remain passive. In the spirit of \citet{cartea2025decentralised}, a natural extension would be to incorporate stochastic liquidity dynamics.

\section*{Acknowledgments}

The authors would like to thank Mohamed Frihat, Vincent Danos, Hamza El Khalloufi and Alejandro Roigé Vázquez for fruitful discussions.

This work used HPC resources from the ``Mésocentre'' computing center of CentraleSupélec and École Normale Supérieure Paris-Saclay supported by CNRS and Région Île-de-France.

\section*{Disclosure of interest}

The authors have no competing interests to declare.

\bibliography{bibliographie.bib}

@article{almgren1999value,
  title={{V}alue under liquidation},
  author={Almgren, Robert and Chriss, Neil},
  journal={Risk},
  volume={12},
  number={12},
  pages={61--63},
  year={1999}
}

@article{almgren2001optimal,
  title={{O}ptimal execution of portfolio transactions},
  author={Almgren, Robert and Chriss, Neil},
  journal={Journal of Risk},
  volume={3},
  pages={5--40},
  year={2001}
}

@article{almgren2003optimal,
  title={{O}ptimal execution with nonlinear impact functions and trading-enhanced risk},
  author={Almgren, Robert F},
  journal={Applied Mathematical Finance},
  volume={10},
  number={1},
  pages={1--18},
  year={2003},
  publisher={Taylor \& Francis}
}

@book{gueant2016financial,
  title={{T}he financial mathematics of market liquidity: {F}rom optimal execution to market making},
  author={Gu{\'e}ant, Olivier},
  volume={33},
  year={2016},
  publisher={CRC Press}
}

@book{cartea2015algorithmic,
  title={{A}lgorithmic and high-frequency trading},
  author={Cartea, {\'A}lvaro and Jaimungal, Sebastian and Penalva, Jos{\'e}},
  year={2015},
  publisher={Cambridge University Press}
}

@article{obizhaeva2013optimal,
  title={{O}ptimal trading strategy and supply/demand dynamics},
  author={Obizhaeva, Anna A and Wang, Jiang},
  journal={Journal of Financial markets},
  volume={16},
  number={1},
  pages={1--32},
  year={2013},
  publisher={Elsevier}
}

@article{alfonsi2008constrained,
  title={{C}onstrained portfolio liquidation in a limit order book model},
  author={Alfonsi, Aur{\'e}lien and Fruth, Antje and Schied, Alexander},
  journal={Banach Center Publications},
  volume={83},
  pages={9--25},
  year={2008}
}

@article{alfonsi2010optimal,
  title={{O}ptimal execution strategies in limit order books with general shape functions},
  author={Alfonsi, Aur{\'e}lien and Fruth, Antje and Schied, Alexander},
  journal={Quantitative Finance},
  volume={10},
  number={2},
  pages={143--157},
  year={2010},
  publisher={Taylor \& Francis}
}

@article{gatheral2012transient,
  title={{T}ransient linear price impact and {F}redholm integral equations},
  author={Gatheral, Jim and Schied, Alexander and Slynko, Alla},
  journal={Mathematical Finance: An International Journal of Mathematics, Statistics and Financial Economics},
  volume={22},
  number={3},
  pages={445--474},
  year={2012},
  publisher={Wiley Online Library}
}

@article{curato2017optimal,
  title={{O}ptimal execution with non-linear transient market impact},
  author={Curato, Gianbiagio and Gatheral, Jim and Lillo, Fabrizio},
  journal={Quantitative Finance},
  volume={17},
  number={1},
  pages={41--54},
  year={2017},
  publisher={Taylor \& Francis}
}

@book{bouchaud2018trades,
  title={{T}rades, quotes and prices: {F}inancial markets under the microscope},
  author={Bouchaud, Jean-Philippe and Bonart, Julius and Donier, Jonathan and Gould, Martin},
  year={2018},
  publisher={Cambridge University Press}
}

@article{busseti2012calibration,
  title={{C}alibration of optimal execution of financial transactions in the presence of transient market impact},
  author={Busseti, Enzo and Lillo, Fabrizio},
  journal={Journal of Statistical Mechanics: Theory and Experiment},
  volume={2012},
  number={09},
  pages={P09010},
  year={2012},
  publisher={IOP Publishing}
}

@article{bochud2007optimal,
  title = {{O}ptimal approximations of power laws with exponentials: {A}pplication to volatility models with long memory},
  author = {Thierry Bochud and Damien Challet},
  journal = {Quantitative Finance},
  volume = {7},
  number = {6},
  pages = {585--589},
  year = {2007},
  publisher = {Routledge},
}

@article{adams2020uniswap,
  title={{U}niswap v2 core},
  author={Adams, Hayden and Zinsmeister, Noah and Robinson, Dan},
  year={2020},
  publisher={Tech. rep., Uniswap},
  note={Available at \url{https://app.uniswap.org/whitepaper.pdf}}
}

@article{adams2021uniswap,
  title={{U}niswap v3 core},
  author={Adams, Hayden and Zinsmeister, Noah and Salem, Moody and Keefer, River and Robinson, Dan},
  year={2021},
  publisher={Tech. rep., Uniswap},
  note={Available at \url{https://app.uniswap.org/whitepaper-v3.pdf}}
}

@article{egorov2019stableswap,
  title={{S}table{S}wap - efficient mechanism for stablecoin liquidity},
  author={Egorov, Michael},
  year={2019},
  publisher={Tech. rep., Curve},
  note={Available at \url{https://classic.curve.finance/files/stableswap-paper.pdf}}
}

@article{martinelli2019non,
  title={{A} non-custodial portfolio manager, liquidity provider, and price sensor},
  author={Martinelli, Fernando and Mushegian, Nikolai},
  publisher={Tech. rep., Balancer},
  year={2019},
  note={Available at \url{https://docs.balancer.fi/whitepaper.pdf}}
}

@article{cartea2025decentralised,
  title={{D}ecentralised finance and automated market making: {E}xecution and speculation},
  author={Cartea, {\'A}lvaro and Drissi, Fay{\c{c}}al and Monga, Marcello},
  journal={Journal of Economic Dynamics and Control},
  volume={177},
  pages={105134},
  year={2025},
  publisher={Elsevier}
}

@book{kirk2004optimal,
  title={Optimal control theory: {A}n introduction},
  author={Kirk, Donald E},
  year={2004},
  publisher={Courier Corporation}
}

@article{tran2024order,
  title={Order book inspired automated market making},
  author={Tran, Tuan and Tran, Duc A and Nguyen, Tam},
  journal={IEEE Access},
  volume={12},
  pages={36743--36763},
  year={2024},
  publisher={IEEE}
}

@book{horn2013matrix,
  title={Matrix analysis},
  author={Horn, Roger A and Johnson, Charles R},
  year={2013},
  edition={2nd},
  publisher={Cambridge University Press}
}

\section*{Appendix}

\appendix
\addtocontents{toc}{\protect\setcounter{tocdepth}{-5}}
\renewcommand*{\thesubsection}{\Alph{subsection}}

\subsection{Proofs}\label{appendix:proofs}

\subsubsection{Proof of Proposition \ref{pr:general_solution}}

\begin{proof}
By introducing a Lagrange multiplier $\lambda \in \mathbb{R}$ for the equality $\sum^{N}_{m=0} \delta_{m} = \xi$ in \eqref{eq:execution_problem_uniswap_v2}, we obtain:
\begin{equation}\label{eq:execution_problem_Lagrange_uniswap_v2}
\begin{aligned}
\delta^{*} = \underset{\delta}{\mathrm{argmax}} \quad & \mathbb{E} \Big [ \sum^{N}_{n=0} \mathcal{C}_{n} \Big ] + \lambda \big ( \xi - \sum^{N}_{n=0} \delta_{n} \big ).
\end{aligned}
\end{equation}

For $m = 0, \ldots, N$, the first-order condition from \eqref{eq:execution_problem_Lagrange_uniswap_v2} reads:
\begin{equation}\label{eq:first_order_condition_uniswap_v2}
\frac{\partial}{\partial \delta_{m}} \mathbb{E} \Big [ \sum^{N}_{n=0} \mathcal{C}_{n} \Big ] = \lambda,
\end{equation}
where,
\begin{equation}\label{eq:partial_derivative_delta_m}
\begin{aligned}
\frac{\partial}{\partial \delta_{m}} \mathbb{E} \Big [ \sum^{N}_{n=0} \mathcal{C}_{n} \Big ] = \mathbb{E} [ f_{m} ] & - \frac{2}{L} \sum^{m}_{n=0} \delta^{*}_{n} \Big ( \sum^{J}_{j=0} \omega_{j} e^{-\rho_{j} (m-n) \Delta} \mathbb{E} [ f_{m} \sqrt{f_{n}} ] \Big ) \\
& - \frac{2}{L} \sum^{N}_{n=m+1} \delta^{*}_{n} \Big ( \sum^{J}_{j=0} \omega_{j} e^{-\rho_{j} (n-m) \Delta} \mathbb{E} [ f_{n} \sqrt{f_{m}} ] \Big ).
\end{aligned}
\end{equation}

We can rewrite the $N+1$ first-order conditions \eqref{eq:first_order_condition_uniswap_v2} in compact matrix form as:
\begin{equation}\label{eq:equation_solution_uniswap_v2}
A \delta^{*} = \frac{L}{2} \big ( B - \lambda \mathbbm{1} \big ),
\end{equation}
where $\mathbbm{1} = \big ( 1, \ldots, 1 \big )^{\top}$. The vector $B \in \mathbb{R}^{N+1}$ and the matrix $A \in \mathbb{R}^{(N+1) \times (N+1)}$, defined in \eqref{eq:vector_B} and \eqref{eq:matrix_A} respectively, are derived from \eqref{eq:partial_derivative_delta_m}. Substituting \eqref{eq:equation_solution_uniswap_v2} in the volume constraint \eqref{eq:volume_constraint} implies:
\begin{equation}
\lambda^{*} = \frac{\mathbbm{1}^{\top} A^{-1} B - \frac{2 \xi}{L}}{\mathbbm{1}^{\top} A^{-1} \mathbbm{1}},
\end{equation}
and plugging $\lambda^{*}$ in \eqref{eq:equation_solution_uniswap_v2} yields the closed-form optimal execution schedule:
\begin{equation}\label{eq:solution}
\delta^{*} = \frac{L}{2} (A^{-1} B - \lambda^{*} A^{-1} \mathbbm{1}).
\end{equation}
\end{proof}

\subsubsection{Proof of Corollary \ref{co:martingale_case}}

\begin{proof}
Under \textnormal{(H2)}, the schedule \eqref{eq:solution} reduces to \eqref{eq:solution_martingale}. It remains to show that the matrix $A$ defined in \eqref{eq:matrix_A} is positive definite, first under \textnormal{(H1)}--\textnormal{(H2)}--\textnormal{(H3)}, then under \textnormal{(H1)}--\textnormal{(H2)}--\textnormal{(H3')}. Once this is established, the objective function is strictly concave, as its Hessian matrix is $-\frac{2}{L} A$; since the volume constraint $\sum_{n=0}^{N} \delta_{n} = \xi$ defines a convex feasible set, the schedule \eqref{eq:solution_martingale} is then the unique maximizer.

We define the symmetric matrices $K^{(j)} \in \mathbb{R}^{(N+1)\times(N+1)}$, for $j = 0, \ldots, J$, and $G \in \mathbb{R}^{(N+1)\times(N+1)}$ by $K^{(j)}_{mn} = e^{-\rho_{j} |m-n| \Delta}$ and $G_{mn} = \mathbb{E} \big [ f_{\max(m,n)} \sqrt{f_{\min(m,n)}} \big ]$. Let $\circ$ denote the element-wise (Hadamard) product; the matrix $A$ admits the following decomposition:
\begin{equation}\label{eq:matrix_A_hadamard_decomposition}
A = \sum^{J}_{j=0} \omega_{j} \big ( K^{(j)} \circ G \big ).
\end{equation}
Since $\omega_{j} > 0$ for $j = 0, \ldots, J$, the matrix $A$ is positive definite once each $K^{(j)} \circ G$ is positive semi-definite and there exists $j$ such that $K^{(j)} \circ G$ is positive definite. By the Schur product theorem \citep[Theorem 7.5.3]{horn2013matrix}, the Hadamard product of two symmetric matrices is positive definite if one is positive definite and the other is positive semi-definite with strictly positive diagonal entries. Here, \textnormal{(H3)} ensures that there exists $j$ such that $\rho_{j} > 0$, so that $K^{(j)}$ is positive definite \citep{alfonsi2008constrained}. It remains to show that $G$ is positive semi-definite with strictly positive diagonal entries.

Let $(e_{i})^{N}_{i=0}$ denote the canonical basis of $\mathbb{R}^{N+1}$ and, for $n = 0, \dots, N$, we define:
\begin{equation}
u_{n} = \sum^{n}_{k=0} \sqrt{\Delta g_{k}} e_{k},
\end{equation}
where $\Delta g_{k} = g_{k} - g_{k-1} \geq 0$ with $g_{k} = \mathbb{E} \big [ f^{3/2}_{k} \big ]$, which is finite under \textnormal{(H1)} and $g_{-1} = 0$. The non-negativity of $\Delta g_{k}$ follows from the conditional Jensen inequality and \textnormal{(H2)}. The matrix $G$ is the corresponding Gram matrix: $\langle u_{m}, u_{n} \rangle = \sum_{i=0}^{\min(m,n)} \Delta g_{i} = g_{\min(m,n)} = G_{mn}$ for $m,n = 0, \dots, N$, where the last equality follows from \textnormal{(H2)} and the tower property. In particular, its diagonal entries are strictly positive given $g_{0} > 0$.

Additionally, let $U$ be the upper triangular matrix whose columns are the vectors $u_{0}, \ldots, u_{N}$:
\begin{equation}
U = \begin{bmatrix}
\sqrt{\Delta g_{0}} & \sqrt{\Delta g_{0}} & \sqrt{\Delta g_{0}} & \cdots & \sqrt{\Delta g_{0}} \\
0 & \sqrt{\Delta g_{1}} & \sqrt{\Delta g_{1}} & \cdots & \sqrt{\Delta g_{1}} \\
0 & 0 & \sqrt{\Delta g_{2}} & \cdots & \sqrt{\Delta g_{2}} \\
\vdots & \vdots & \vdots & \ddots & \vdots \\
0 & 0 & 0 & \cdots & \sqrt{\Delta g_{N}}
\end{bmatrix}.
\end{equation}
Then $G = U^{\top} U$. Consequently, for all non-zero $x \in \mathbb{R}^{N+1}$, $x^{\top} G x = \|Ux\|^{2} \geq 0$, which shows that $G$ is positive semi-definite.

When the impact is fully permanent, \textnormal{(H3)} no longer holds and the roles of $K^{(j)}$ and $G$ in the Schur product theorem can be interchanged; assumption \textnormal{(H3')} is then required to ensure $\Delta g_{k} > 0$, which in turn implies that the matrix $G$ is positive definite.
\end{proof}

\subsubsection{Proof of Corollary \ref{co:GBM}}

\begin{proof}
Under the geometric Brownian motion assumption \eqref{eq:Geometric_Brownian_motion}, the expectations appearing in \eqref{eq:vector_B} and \eqref{eq:matrix_A} admit closed-form expressions. For $m = 0, \ldots, N$, the expected fundamental price reads $\mathbb{E} [ f_{m} ] = f_{0} e^{\mu m \Delta}$. Moreover, for $n \leq m$, we have:
\begin{equation}\label{eq:E_price_lognormal}
\mathbb{E} [ f_{m} \sqrt{f_{n}} ] = f_{0} \sqrt{f_{0}} e^{\mu (m + \frac{n}{2}) \Delta + \frac{3\sigma^{2}}{8} n \Delta},
\end{equation}
and symmetrically for $n > m$.

We start with a single resilience factor (i.e., $J = 0$), for which the matrix $A$ from \eqref{eq:matrix_A} takes the following form, up to the multiplicative factor $f_{0} \sqrt{f_{0}}$, which is omitted for clarity:
\begin{equation}
A =
\begin{bmatrix}
b_0 & a b_0 & a^2 b_0 & a^3 b_0 & \cdots & a^N b_0 \\
a b_0 & b_1 & a b_1 & a^2 b_1 & \cdots & a^{N-1} b_1 \\
a^2 b_0 & a b_1 & b_2 & a b_2 & \cdots & a^{N-2} b_2 \\
a^3 b_0 & a^2 b_1 & a b_2 & b_3 & \cdots & a^{N-3} b_3 \\
\vdots & \vdots & \vdots & \vdots & \ddots & \vdots \\
a^N b_0 & a^{N-1} b_1 & a^{N-2} b_2 & a^{N-3} b_3 & \cdots & b_N
\end{bmatrix}.
\end{equation}
where $a = e^{- (\rho - \mu) \Delta}$ and $b_{n} = e^{\big ( \frac{3 \mu}{2} + \frac{3 \sigma^{2}}{8} \big ) n \Delta}$. Let $(e_{i})^{N}_{i=0}$ denote the canonical basis of $\mathbb{R}^{N+1}$ and, in the spirit of \citet{alfonsi2008constrained}, we define the vectors $v_{0}, \ldots, v_{N}$ by the recursive formula:
\begin{equation}\label{eq:recursion_single}
\left\{
    \begin{array}{ll}
        v_{0} = e_{0} \\
        v_{n} = v_{n-1} a + e_{n} \sqrt{\gamma_{n}},
    \end{array}
\right.
\end{equation}
where $\gamma_{n} = b_{n} - a^{2} b_{n-1} > 0$ for $n = 1, \ldots, N$, with $\gamma_{0} = 1$, the strict positivity following from the drift condition $\mu < \frac{3 \sigma^{2}}{4} + 4 \rho$. The matrix $A$ is the corresponding Gram matrix: $\langle v_{m}, v_{n} \rangle = a^{|m-n|} b_{\min (m,n)} = A_{mn}$ for $m,n = 0, \ldots, N$. Indeed, first by induction: $\langle v_{n}, v_{n} \rangle = b_{n}$. Second, for $n < m$ we have: $\langle v_{m}, v_{n} \rangle = a^{m-n} \langle v_{n}, v_{n} \rangle$ and for $m < n$: $\langle v_{m}, v_{n} \rangle = a^{n-m} \langle v_{m}, v_{m} \rangle$. From the recursive formula \eqref{eq:recursion_single}, we also have:
\begin{equation}
v_{n} = \sum^{n}_{k=0} \big [ a^{n-k} \sqrt{\gamma_{k}} \big ] e_{k}.
\end{equation}
Let $V$ be the upper triangular matrix whose columns are the vectors $v_0, \ldots, v_N$:
\begin{equation}
V = 
\begin{bmatrix}
1 & a & a^{2} & \cdots & a^{N} \\
0 & \sqrt{\gamma_{1}} & a \sqrt{\gamma_{1}} & \cdots & a^{N-1} \sqrt{\gamma_{1}} \\
0 & 0 & \sqrt{\gamma_{2}} & \cdots & a^{N-2} \sqrt{\gamma_{2}} \\
\vdots & \vdots & \vdots & \ddots & \vdots \\
0 & 0 & 0 & \cdots & \sqrt{\gamma_{N}}
\end{bmatrix}.
\end{equation}

Then, $A = V^{\top} V$. Indeed, for $m, n = 0, \ldots, N$:
\begin{equation}
(V^{\top} V)_{mn} = \sum_{k=0}^N V_{km} V_{kn} = \sum_{k=0}^{\min(m,n)} V_{km} V_{kn},
\end{equation}
since $V$ is upper triangular. Substituting the expression for $V_{km}$ and $V_{kn}$:
\begin{equation}
(V^{\top} V)_{mn} = \sum_{k=0}^{\min(m,n)} \left[a^{m-k} \sqrt{\gamma_{k}} \right] \left[a^{n-k} \sqrt{\gamma_{k}} \right]
= a^{m+n} \sum_{k=0}^{\min(m,n)} a^{-2k} \big ( b_{k} - a^{2} b_{k-1} \big ),
\end{equation}
where $b_{-1} = 0$. This sum telescopes to $a^{-2 \min(m,n)} b_{\min(m,n)}$, yielding $(V^{\top} V)_{mn} = a^{|m-n|} b_{\min(m,n)} = A_{mn}$. Under the drift condition, $V$ is invertible, and therefore $A$ is positive definite. As in the proof of Corollary \ref{co:martingale_case}, this guarantees that the optimal strategy $\delta^{*}$ in \eqref{eq:solution_reformulated} is the unique maximizer of the execution problem \eqref{eq:execution_problem_uniswap_v2}. Additionally, $A^{-1} = V^{-1} (V^{-1})^{\top}$, and the inverse matrix $A^{-1}$ is tridiagonal, given by \eqref{eq:inverse_matrix_A}.

Conversely, if $A$ is positive definite, then by Sylvester's criterion \citep[Theorem 7.2.5]{horn2013matrix}, all leading principal minors of $A$ are strictly positive, that is, $\det(A^{(n)}) > 0$ for all $n = 0, \ldots, N$, where $A^{(n)}$ denotes the leading principal submatrix of order $n+1$. By induction using the Laplace expansion along the last row, we obtain:
\begin{equation}\label{eq:minors}
\det(A^{(n)}) = (f_{0} \sqrt{f_{0}})^{n+1} \prod_{k=0}^{n} \gamma_{k}.
\end{equation}
Since $f_{0} \sqrt{f_{0}} > 0$, this implies $\mu < \frac{3\sigma^{2}}{4} + 4\rho$, which is therefore necessary and sufficient for the positive definiteness of $A$.

In the general multi-kernel case ($J \geq 1$), the condition $\mu < \frac{3\sigma^{2}}{4} + 4\min_{j}\rho_{j}$ guarantees the positive definiteness of $A$ via the decomposition \eqref{eq:matrix_A_hadamard_decomposition} and therefore the uniqueness of $\delta^{*}$. This condition is sufficient but not necessary for $J \geq 1$.
\end{proof}

\subsubsection{Proof of Proposition \ref{pr:dynamic_programming_closed_loop}}\label{appendix:Proof_Bellman_Uniswap_v2}

\begin{proof}
We consider the $\textit{ansatz}$ \eqref{eq:value_function_closed_loop_uniswap_v2} with coefficients $A_{n}, (B^{j}_{n})^{J}_{j=0}, C_{n}, D_{n}, (E_{n})^{J}_{j=0}$ and $(F_{n})^{J}_{j_{1},j_{2}=0}$ determined by backward recursion. Substituting \eqref{eq:value_function_closed_loop_uniswap_v2} into the Bellman equation \eqref{eq:Bellman_equation_closed_loop_uniswap_v2} yields a quadratic objective in $\delta_{n}$:
\begin{equation}
\begin{aligned}
v_{n}(x_{n}, (I^{j}_{n})^{J}_{j=0}, f_{n}) = \sup_{\delta_{n}} \ & - \delta^{2}_{n} f_{n} \sqrt{f_{n}} \Big [ \frac{1}{L} + \frac{2}{L} \sum^{J}_{j=0} B^{j}_{n+1} e^{(\mu-\rho_{j}) \Delta} - C_{n+1} e^{(\frac{3 \mu}{2} + \frac{3 \sigma^{2}}{8}) \Delta} \\
& - \frac{4}{L^{2}} \sum^{J}_{j_{1}=0} \sum^{J}_{j_{2}=0} F^{j_{1}, j_{2}}_{n+1} e^{(\frac{\mu}{2} - \rho_{j_{1}} - \rho_{j_{2}} - \frac{\sigma^{2}}{8}) \Delta} \Big ] \\
& + \delta_{n} \Big [ f_{n} \big ( 1 - A_{n+1} e^{\mu \Delta} + \frac{2}{L} \sum^{J}_{j=0} E^{j}_{n+1} e^{(\frac{\mu}{2} - \rho_{j} - \frac{\sigma^{2}}{8}) \Delta} \big ) \\
& + f_{n} \sum^{J}_{j_{1}=0} I^{j_{1}}_{n} \big ( - \omega_{j_{1}} - B^{j_{1}}_{n+1} e^{(\mu - \rho_{j_{1}}) \Delta} + \frac{4}{L} \sum^{J}_{j_{2}=0} F^{j_{1}, j_{2}}_{n+1}  e^{(\frac{\mu}{2} - \rho_{j_{1}} - \rho_{j_{2}} - \frac{\sigma^{2}}{8}) \Delta} \big ) \\
& + x_{n} f_{n} \sqrt{f_{n}} \big ( \frac{2}{L} \sum^{J}_{j=0} B^{j}_{n+1} e^{(\mu - \rho_{j}) \Delta} - 2 C_{n+1} e^{(\frac{3\mu}{2} + \frac{3\sigma^{2}}{8}) \Delta} \big ) \Big ] \\
& + x_{n} f_{n} \Big [ A_{n+1} e^{\mu \Delta} + \sum^{J}_{j=0} B^{j}_{n+1} e^{(\mu - \rho_{j}) \Delta} I^{j}_{n} + C_{n+1} e^{(\frac{3 \mu}{2} + \frac{3 \sigma^{2}}{8}) \Delta} x_{n} \sqrt{f_{n}} \Big ] \\
& + \sqrt{f_{n}} \Big [ D_{n+1} e^{(\frac{\mu}{2} - \frac{\sigma^{2}}{8}) \Delta} + \sum^{J}_{j=0} E^{j}_{n+1} e^{(\frac{\mu}{2} - \rho_{j} - \frac{\sigma^{2}}{8}) \Delta} I^{j}_{n} \\
& + \sum^{J}_{j_{1}=0} \sum^{J}_{j_{2}=0} F^{j_{1}, j_{2}}_{n+1} e^{(\frac{\mu}{2} - \rho_{j_{1}} - \rho_{j_{2}} - \frac{\sigma^{2}}{8}) \Delta} I^{j_{1}}_{n} I^{j_{2}}_{n} \Big ],
\end{aligned}
\end{equation}
and leads to the explicit control \eqref{eq:solution_closed_loop_uniswap_v2}. The substitution of the optimal control into the Bellman equation \eqref{eq:Bellman_equation_closed_loop_uniswap_v2} and the terminal condition \eqref{eq:Bellman_equation_terminal_closed_loop_uniswap_v2}, and identification of coefficients in the polynomial basis yield the backward recursions \eqref{eq:parameters_closed_loop_uniswap_v2}.
\end{proof}

\subsection{Optimal execution on Uniswap v2: the two-period martingale case}\label{appendix:two_period}

In the case $\mu = 0$, $N=1$ and $J=0$, the matrix $A$ defined in \eqref{eq:matrix_A} reads:
\begin{equation}
A = f_{0} \sqrt{f_{0}}
\begin{bmatrix}
1 & e^{-\rho \Delta} \\
e^{-\rho \Delta} & e^{\frac{3\sigma^{2}}{8}\Delta}
\end{bmatrix}.
\end{equation}
The corresponding optimal trades are:
\begin{equation}
\delta^{*}_{0} = \xi \frac{e^{\frac{3\sigma^{2}}{8} \Delta} - e^{- \rho \Delta}}{e^{\frac{3\sigma^{2}}{8} \Delta} + 1 - 2 e^{- \rho \Delta}}, \ \delta^{*}_{1} = \xi \frac{1 - e^{- \rho \Delta}}{e^{\frac{3\sigma^{2}}{8} \Delta} + 1 - 2 e^{- \rho \Delta}}.
\end{equation}
A direct comparison yields $\delta^{*}_{0} > \delta^{*}_{1}$, which establishes that the first trade is larger than the last one in the two-period framework, consistently with the numerical observations reported in Section \ref{section:illustrations}.

Differentiating with respect to volatility $\sigma$ gives:
\begin{equation}
\frac{\partial \delta^{*}_{0}}{\partial \sigma} = \xi \frac{3 \sigma}{4} \Delta  e^{\frac{3\sigma^{2}}{8} \Delta} \frac{1 - e^{- \rho \Delta}}{\big ( e^{\frac{3\sigma^{2}}{8} \Delta} + 1 - 2 e^{- \rho \Delta} \big )^{2}} > 0,
\end{equation}
showing that higher volatility increases the size of the initial trade. Differentiation with respect to the resilience parameter $\rho$ gives:
\begin{equation}
\frac{\partial \delta^{*}_{0}}{\partial \rho} = \xi \Delta  e^{- \rho \Delta} \frac{1 - e^{\frac{3\sigma^{2}}{8} \Delta}}{\big ( e^{\frac{3\sigma^{2}}{8} \Delta} + 1 - 2 e^{- \rho \Delta} \big )^{2}} < 0,
\end{equation}
which indicates that faster price recovery of the liquidity pool reduces the size of the initial trade.

Finally, we study the asymptotic behavior of the optimal trades. When volatility becomes large, the solution converges to a fully front-loaded strategy:
\begin{equation}
\lim_{\sigma \rightarrow + \infty} \delta^{*}_{0} = \xi, \ \lim_{\sigma \rightarrow + \infty} \delta^{*}_{1} = 0.
\end{equation}
When the resilience parameter becomes large, we obtain:
\begin{equation}
\lim_{\rho \rightarrow + \infty} \delta^{*}_{0} = \xi \frac{e^{\frac{3\sigma^{2}}{8} \Delta}}{e^{\frac{3\sigma^{2}}{8} \Delta} + 1}, \ \lim_{\rho \rightarrow + \infty} \delta^{*}_{1} = \xi \frac{1}{e^{\frac{3\sigma^{2}}{8} \Delta} + 1}.
\end{equation}
In this regime, the price recovers instantaneously, and the trade allocation reflects only the effects of volatility, with a persistent asymmetry favoring earlier execution under higher expected price impact.

\subsection{Backward algorithm}\label{appendix:algorithm}

Under the two-layer liquidity framework introduced in Section \ref{section:two_layers}, the backward algorithm is provided in the form of pseudocode in Algorithm \ref{alg:grid_based_solver}.
\begin{algorithm}
\caption{Two-layer liquidity backward algorithm}
\label{alg:grid_based_solver}
\begin{algorithmic}

\Require \\
\begin{itemize}
\renewcommand{\labelitemi}{\texttt{-}}
\item The strategy parameters: $T, N$ and $\xi$
\item The market parameters: $f_{0}, \overline{p}, L_{0}$ and $L_{1}$
\item The model parameters: $J, (\rho_{j})^{J}_{j=0}, (\omega_{j})^{J}_{j=0}, \mu$ and $\sigma$
\item The grid parameters: $K_{x}, K_{I}, K_{f}$ and $z$
\end{itemize}

\Ensure \\
\begin{itemize}
\renewcommand{\labelitemi}{\texttt{-}}
\item The estimated optimal execution strategy $\delta^{*}$
\item The associated value function $v$
\end{itemize}

\medskip

\State Set the grids according to \eqref{eq:grid_p}, \eqref{eq:grid_x}, \eqref{eq:grid_i} and \eqref{eq:grid_y}

\State Set $v_{N}$ according to \eqref{eq:Bellman_equation_closed_loop_uniswap_v3} \Comment{The terminal value function is initialized.}

\medskip

\For{$n = N-1, \ldots, 0$} \Comment{Backward iteration.}
    \For{$k_{f} = 0, \ldots, K_{f}$, $k_{x} = 0, \ldots, K_{x}$, $k^{j}_{i} = 0, \ldots, K_{I}$, $j = 0, \ldots, J$}
        \For{$k = 0, \ldots, k_{x}$}
            \smallskip

            \State Set $\delta_{n} = x^{(k)}$ \Comment{Candidate trade from the inventory grid.}
            \medskip

            \State Compute the resulting cash-flow $\mathcal{C}_{n}(x^{(k_{x})}, (I^{(k^{j}_{i})})^{J}_{j=0}, f^{(k_{f})}, \delta_{n})$ from \eqref{eq:cash_tn_v3}
            \medskip

            \State Update the cumulative price impacts $I^{j}_{n+1} (x^{(k_{x})}, (I^{(k^{j}_{i})})^{J}_{j=0}, f^{(k_{f})}, \delta_{n})$ from \eqref{eq:dynamic_cumulative_impact_closed_loop_v3}
            \medskip

            \State Update the remaining inventory $x_{n+1} = x^{(k_{x})} - \delta_{n}$
            \medskip

            \State Perform a ($J+1$)-d linear interpolation to derive $v_{n+1} (x^{(k_{x})}, (I^{(k^{j}_{i})})^{J}_{j=0}, f^{(k_{f})})$ for $k_{f} = 0, \ldots, K_{f}$
            \medskip

            \State Estimate the conditional expectation $\overline{v}_{n+1}$ using \eqref{eq:cond_expectation}
            \medskip

            \State Store the total value $\mathcal{C}_{n} + \overline{v}_{n+1}$ for each candidate trade $\delta_{n}$
            \medskip

        \EndFor
        \smallskip

            \State Determine the optimal trade:
            $$\delta^{*}_{n} (x^{(k_{x})}, (I^{(k^{j}_{i})})^{J}_{j=0}, f^{(k_{f})}) = \underset{\delta_{n}}{\mathrm{argmax}} \ \mathcal{C}_{n} + \overline{v}_{n+1}$$

            \State Update the value function $v_{n} (x^{(k_{x})}, (I^{(k^{j}_{i})})^{J}_{j=0}, f^{(k_{f})})$ accordingly
            \medskip

    \EndFor
\EndFor

\end{algorithmic}
\end{algorithm}

\subsection{Consistency and stability checks}\label{appendix:checks}

\begin{figure}
    \centering
    \begin{subfigure}{\linewidth} 
        \centering
        \includegraphics[scale = 0.6]{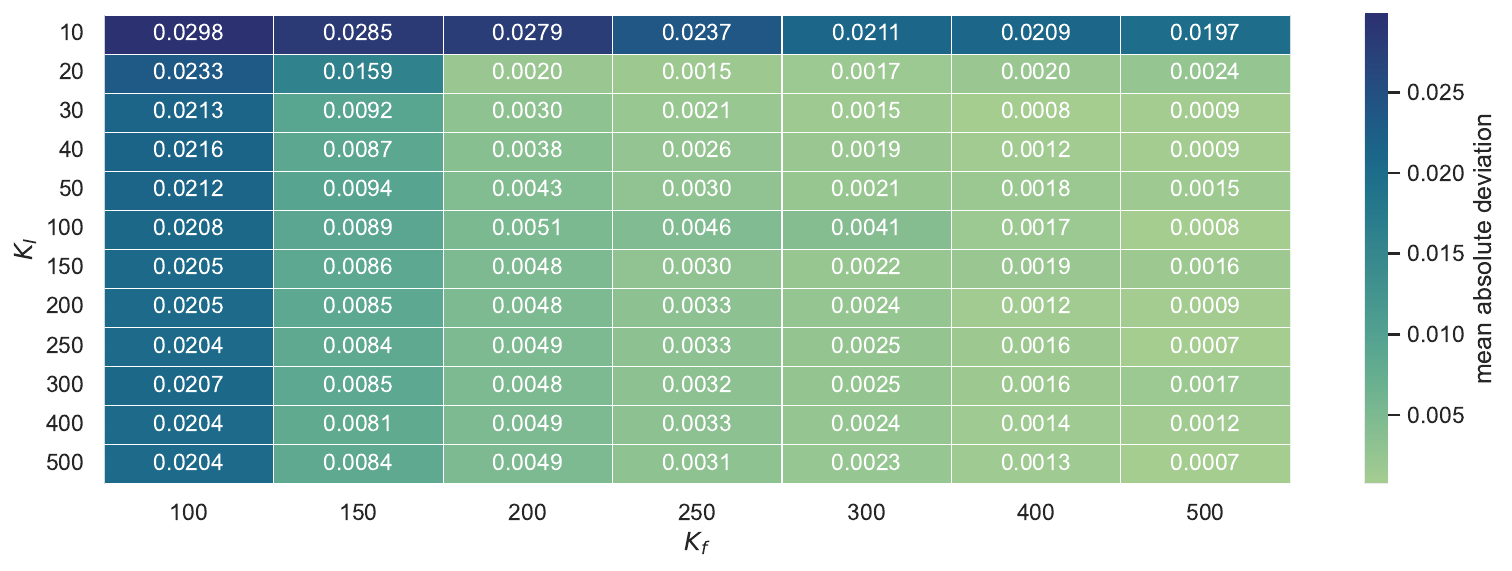}
        \caption{Mean absolute error}
        \label{fig:checks_v2_a}
    \end{subfigure}
    \vfill
    \medskip
    \begin{subfigure}{\linewidth}
        \centering
        \includegraphics[scale = 0.6]{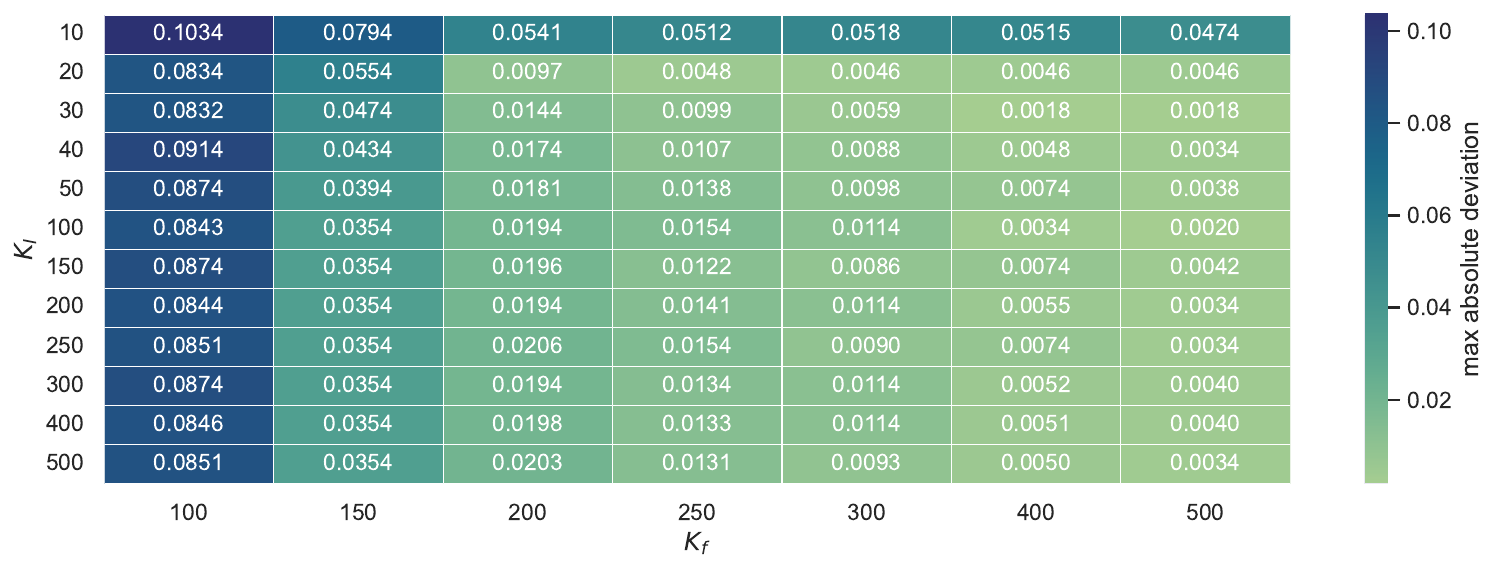}
        \caption{Maximum absolute error}
        \label{fig:checks_v2_b}
    \end{subfigure}
    \caption{Mean and maximum absolute error with respect to $K_{I}$ and $K_{f}$ for $K_{x} = 250$ in the Uniswap v2 framework; $T = 1$, $N = 10$, $\xi = 1$, $L = 1000$, $f_{0} = 1$, $\mu = 0$, $\sigma = 0.3$, $J = 0$ and $\rho = 3$.}
    \label{fig:checks_v2}
\end{figure}
First, we present consistency checks conducted within the Uniswap v2 framework. We compare the numerical results obtained using the algorithm introduced in Section \ref{section:algorithm} when the liquidity layers coincide (i.e., $L_{0} = L_{1}$) with the closed-form solution derived in Section \ref{section:dynamic_programming_closed_loop}, both evaluated along the mean price trajectory. Specifically, we fix $K_{x} = 250$ and compute the maximum and mean absolute errors for different values of the grid sizes $K_{f}$ and $K_{I}$. Results are reported in Figures \ref{fig:checks_v2_a}, \ref{fig:checks_v2_b} and demonstrate that the discretization scheme provides accurate estimates of the optimal strategy in the Uniswap v2 framework. These results also indicate that the accuracy of the numerical solution is more sensitive to the price grid size than to that of the cumulative market impact. Indeed, once $K_{I}$ exceeds $30$, both mean and maximum errors remain nearly unchanged for fixed values of $K_{f}$.

\begin{figure}
    \centering
    \begin{subfigure}{\linewidth} 
        \centering
        \includegraphics[scale = 0.6]{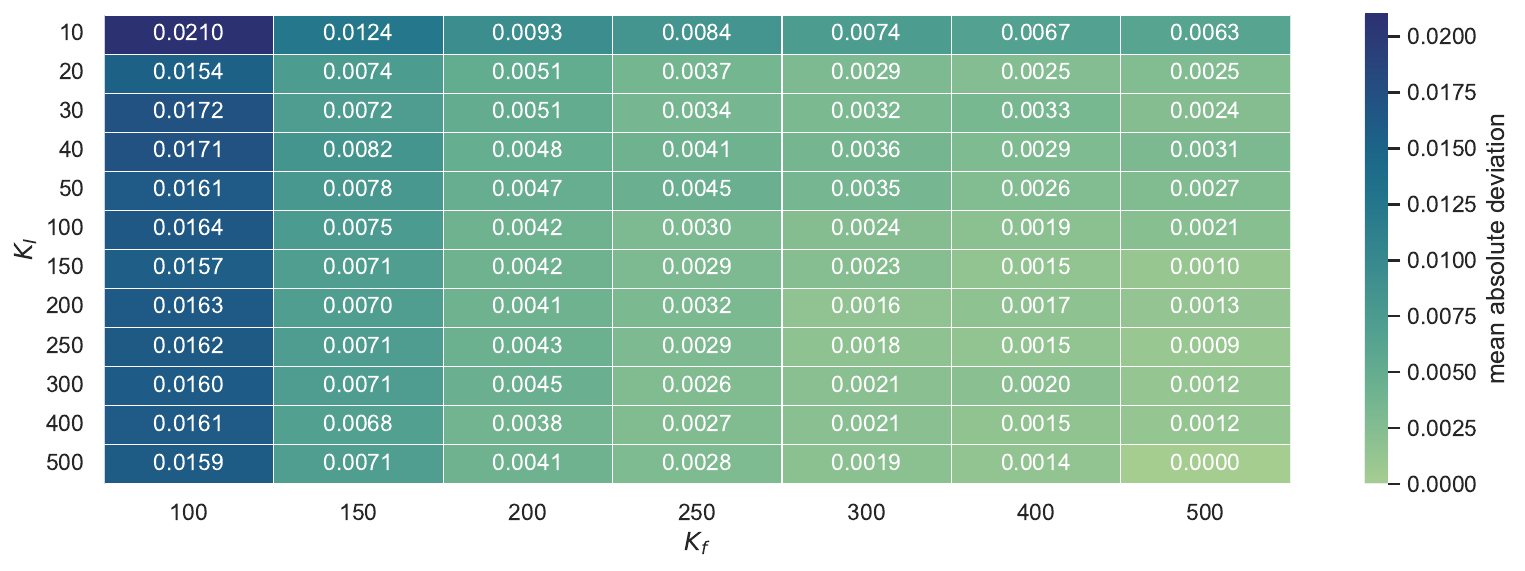}
        \caption{Mean absolute deviation}
        \label{fig:checks_v3_a}
    \end{subfigure}
    \vfill
    \medskip
    \begin{subfigure}{\linewidth}
        \centering
        \includegraphics[scale = 0.6]{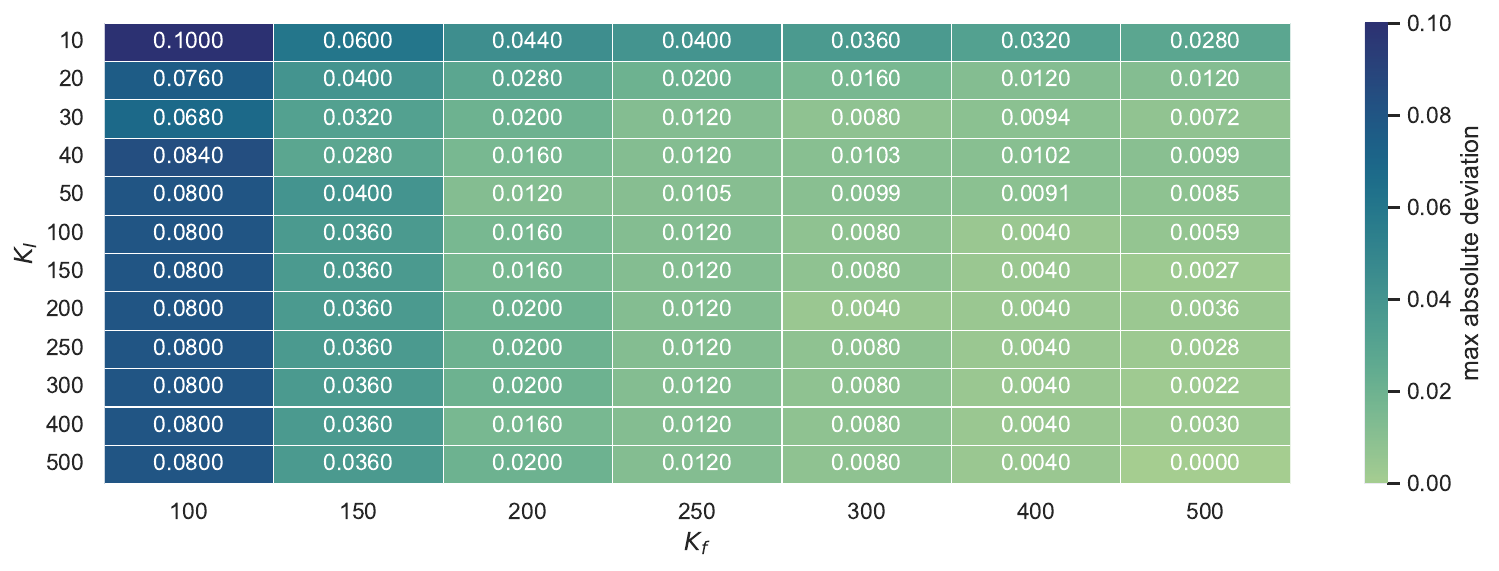}
        \caption{Maximum absolute deviation}
        \label{fig:checks_v3_b}
    \end{subfigure}
    \caption{Mean and maximum absolute deviation with respect to $K_{I}$ and $K_{f}$ for $K_{x} = 250$ in the two-layer liquidity framework; $T = 1$, $N = 10$, $\xi = 1$, $L_{0} = 1000$, $L_{1} = 500$, $f_{0} = 1$, $\overline{s} = -25$bps, $\mu = 0$, $\sigma = 0.3$, $J = 0$ and $\rho = 3$.}
    \label{fig:checks_v3}
\end{figure}
Second, we assess the stability of the numerical scheme within the two-layer liquidity framework introduced in Section \ref{section:two_layers} by varying the grid resolution. We fix the threshold level at $\overline{s} = -25$bps and set $K_{x} = 250$. We compute the maximum and mean absolute deviations relative to a reference solution obtained with the high-resolution configuration ($K_{f} = 500$, $K_{x} = 250$, and $K_{I} = 500$), presented in Section \ref{section:numerical_results_v3}. The results, reported in Figures \ref{fig:checks_v3_a}, \ref{fig:checks_v3_b}, show that the deviation consistently decreases as the grid resolution increases. Moreover, as in the Uniswap v2 setting, accuracy is more sensitive to the resolution of the price grid than to that of the cumulative impact grid. Finally, the computational cost of the high-resolution configuration is significant, requiring approximately $45$ minutes. However, with a coarser grid of $K_{f} = 250$, $K_{x} = 250$, and $K_{I} = 50$, the computation time drops to $2$ minutes, with a maximum deviation of $0.0105$ and a mean deviation of $0.0045$.

\subsection{Beyond the linear approximation}\label{appendix:exact_dynamic_programming}

We now consider the closed-loop dynamic programming framework of Section \ref{section:dynamic_programming_closed_loop} with the exact cash-flow and price impact formulas from the CPAMM pricing rule, instead of their first-order approximations. The remaining inventory still evolves according to \eqref{eq:dynamic_inventory}. The dynamics of the cumulative price impacts are derived from the exact 
post-swap spot price \eqref{eq:post_swap_spot_price} and read:
\begin{equation}\label{eq:dynamic_cumulative_impact_exact}
\left\{
    \begin{array}{ll}
        I^{\text{exact},j}_{0} = 0 \\
        I^{\text{exact},j}_{n+1} = e^{-\rho_{j} \Delta} \Big ( I^{\text{exact},j}_{n} + 1 - \big (1 + \frac{\delta_{n} \sqrt{p_{n}}}{L} \big )^{-2} \Big ),
    \end{array}
\right.
\end{equation}
for $j = 0, \ldots, J$, where $p_{n} = f_{n} \big ( 1 - \sum^{J}_{j=0} \omega_{j} I^{\text{exact},j}_{n} \big )$ denotes the spot price at time $t_{n}$. The exact relative price impact $1 - \big (1 + \frac{\delta_{n}\sqrt{p_{n}}}{L} \big )^{-2}$ is used instead of the first-order approximation $\frac{2\delta_{n}\sqrt{f_{n}}}{L}$. The cash-flow at time $t_{n}$, from \eqref{eq:formula_delta_b}, is:
\begin{equation}\label{eq:cash_exact}
\mathcal{C}^{\text{exact}}_{n} = \frac{\delta_{n} \, p_{n}}{1 + \frac{\delta_{n} \sqrt{p_{n}}}{L}}.
\end{equation}

The associated Bellman equation reads:
\begin{equation}\label{eq:Bellman_exact}
v^{\text{exact}}_{n}(x_{n}, (I^{\text{exact},j}_{n})^{J}_{j=0}, f_{n}) = \sup_{\delta_{n}} \ \mathcal{C}^{\text{exact}}_{n} + \mathbb{E} \Big [ v^{\text{exact}}_{n+1} \big(x_{n+1}, (I^{\text{exact},j}_{n+1})^{J}_{j=0}, f_{n+1} \big) | f_{n} \Big ],
\end{equation}
with terminal condition:
\begin{equation}\label{eq:terminal_exact}
v^{\text{exact}}_{N}(x_{N}, (I^{\text{exact},j}_{N})^{J}_{j=0}, f_{N}) = \frac{x_{N} \, p_{N}}{1 + \frac{x_{N} \sqrt{p_{N}}}{L}}.
\end{equation}

The nonlinearity of \eqref{eq:dynamic_cumulative_impact_exact} and \eqref{eq:cash_exact} prevents closed-form expressions for the value function and the optimal control. The problem is therefore solved numerically by adapting Algorithm \ref{alg:grid_based_solver}, with the price impact, cash-flow, and terminal condition formulas \eqref{eq:dynamic_cumulative_impact_closed_loop_v3}, \eqref{eq:cash_tn_v3} and \eqref{eq:Bellman_equation_closed_loop_uniswap_v3} replaced by their exact counterparts \eqref{eq:dynamic_cumulative_impact_exact}, \eqref{eq:cash_exact} and \eqref{eq:terminal_exact}. This in turn requires adjusting the grid associated with the cumulative impact induced by the $j$-th resilience factor to account for the exact price impact:
\begin{equation}\label{eq:grid_i}
I^{(k^{j}_{i})} = e^{-\rho_{j} \Delta} \Big ( 1 - \big (1 + \frac{\xi \sqrt{f^{(K_{f})}}}{L} \big )^{-2} \Big ) \frac{k^{j}_{i}}{K_{I}},
\end{equation}
for $k^{j}_{i} \in \{ 0, \ldots, K_{I} \}$; the remainder of Algorithm \ref{alg:grid_based_solver} is unchanged.

We compare the exact solution of this section with the first-order one of Section \ref{section:dynamic_programming_closed_loop}, both approximated numerically using Algorithm \ref{alg:grid_based_solver} with the same parameter set as in the Uniswap v2 case of Appendix \ref{appendix:checks} ($T = 1$, $N = 10$, $\xi = 1$, $L = 1000$, $f_{0} = 1$, $\mu = 0$, $\sigma = 0.3$, $J = 0$ and $\rho = 3$). With the high-resolution configuration ($K_{f} = 500$, $K_{x} = 250$, and $K_{I} = 500$), the maximum and mean absolute deviations between the two solutions are $0.0007$ and $0.0001$, respectively; with the coarser grid ($K_{f} = 250$, $K_{x} = 250$, and $K_{I} = 50$), both deviations are even lower. These deviations are negligible, confirming the validity of the first-order approximation as long as trades are small relative to the reserve of the token being sold, which is the case for the set of parameters considered.

\end{document}